\documentclass[%
 reprint,
 amsmath,amssymb,
 aps,
]{revtex4-2}
\usepackage{xcolor}
\usepackage{graphicx}% Include figure files
\usepackage{dcolumn}% Align table columns on decimal point
\usepackage{bm}% bold math
\usepackage{circuitikz}
\usepackage{float}
\usepackage{mleftright}
\usepackage[normalem]{ulem}
\begin{document}

\preprint{APS/123-QED}

\title{Mitigating transients in flux-control signals in a superconducting quantum processor }

\author{Anuj Aggarwal}
\author{Jorge Fern\'andez-Pend\'as}
\author{Tahereh Abad}
\author{Daryoush Shiri}
\author{Halld\'{o}r Jakobsson}
\author{Marcus Rommel}
\author{Andreas Nylander}
\author{Emil Hogedal}
\author{Amr Osman}
\author{Janka Bizn\'{a}rov\'{a}}
\author{Robert Rehammar}
\author{Michele Faucci Giannelli}
\author{Anita Fadavi Roudsari}
\author{Jonas Bylander}
\author{Giovanna Tancredi}
  \email[Contact author: ]{tancredi@chalmers.se}

\affiliation{Department of Microtechnology and Nanoscience, 
Chalmers University of Technology, SE-412 96 Gothenburg, Sweden}

\date{\today}% It is always \today, today,

\begin{abstract}
Magnetic flux-tunable qubits and couplers are essential components of superconducting quantum processors. However, it is challenging to dynamically control these elements with high precision via current pulses because of signal distortions and transients. In particular, transients can persist at times longer than control pulses, adversely affecting subsequent control operations. We model the flux control line as a first-order RC circuit and introduce a class of pulses specifically designed to suppress long-time transients. We theoretically demonstrate the robustness of these pulses against parameter mischaracterization and provide experimental evidence of their effectiveness in mitigating transients when applied to a flux-tunable element. The proposed pulse design offers a practical solution for mitigating long-time transients, enabling efficient and reliable experimental tune-up without requiring detailed flux-line characterization.
\end{abstract}

\maketitle

\section{Introduction}
Superconducting quantum processors have emerged as one of the leading platforms for the physical realization of quantum computers because of their scalability and performance~\cite{acharya2024quantum, krinner2022realizing,jurcevic2021demonstration,kosen2022building,marques2022logical,zhao2022realization}. The majority of superconducting processors rely on flux-tunable circuit elements, whose transition frequency is varied statically and/or dynamically to implement two-qubit gates~\cite{foxen2020demonstrating, rol2019fast, marxer2023long, li2024realization,xu2020high}, reset operations~\cite{reed2010fast, chen2024fast}, and leakage reduction schemes~\cite{yang2024coupler,mcewen2021removing,lacroix2023fast}. Such an element is termed a qubit or a coupler depending on its role in the processor, but in both cases its frequency is controlled in the same way: inductively, via a dedicated flux-control line. Accurate flux control is therefore essential for high-fidelity quantum operations.

\begin{figure*}
    \centering
    \includegraphics[scale=0.4 ]{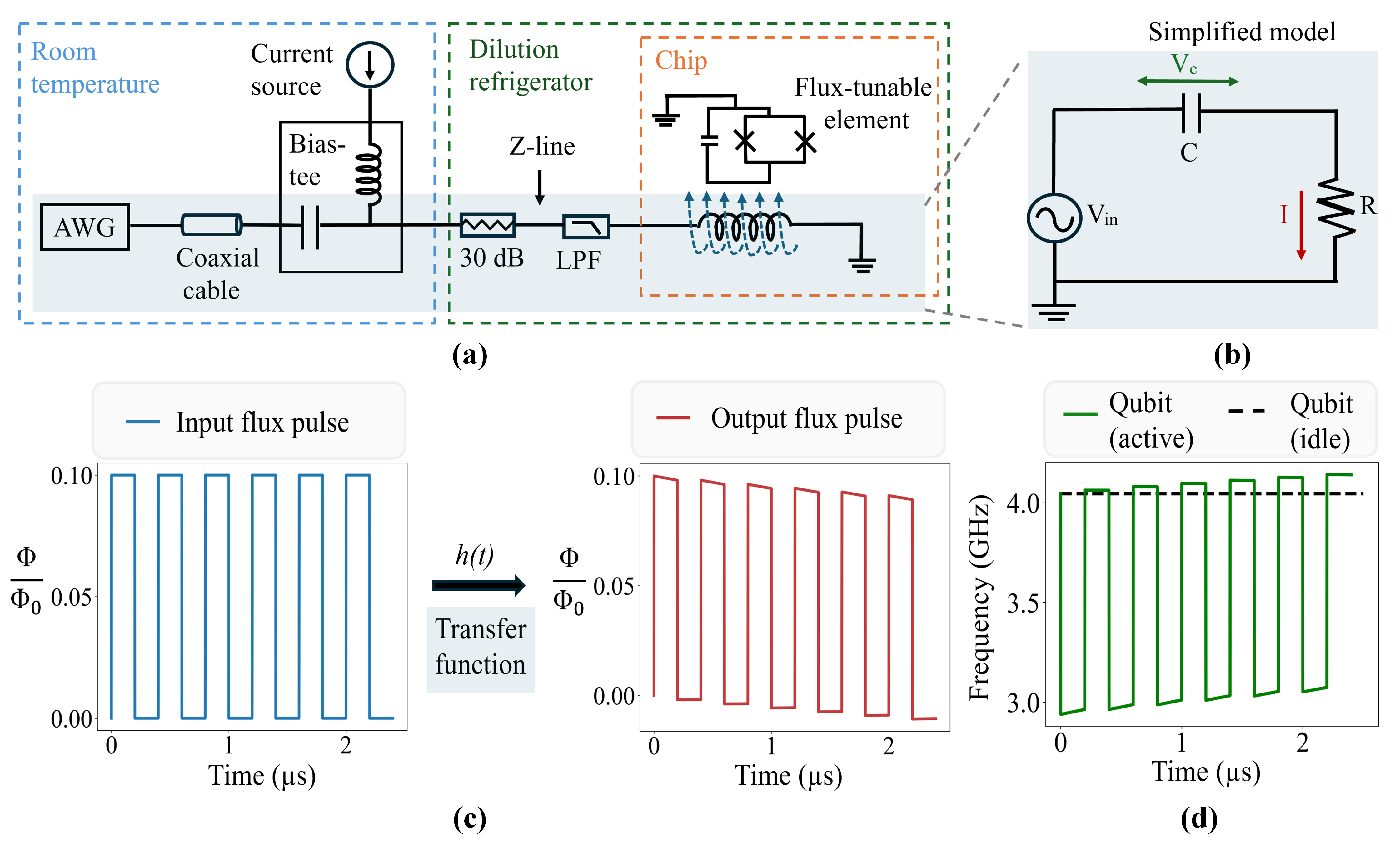}
    \caption{ (a)  Circuit schematic of a typical flux-control  (Z) line. An arbitrary waveform generator (AWG) sends pulses via a bias-tee. The attenuators and low-pass filter reduce noise. The flux line is terminated on chip and is used to control the frequency of a flux-tunable circuit element, i.e. a qubit or a coupler, depending on its role in the processor. (b) Simplified first-order RC model of the flux-control line. The bias tee constitutes the dominant contribution to the capacitance C while attenuators contribute to resistance R. (c) Example demonstration of long-timescale transients using the simplified model, showing how the input pulse is deformed by the transfer function $h(t)$ of the transmission line, with decay time constant $\tau = 10~\text{\textmu s}$. (d) Illustration of the impact of transients on the frequency of a tunable qubit, assuming $\omega_{max}/2\pi=5~\mathrm{GHz}$ and an idle flux bias  $\Phi=0.2~\mathrm{\Phi_0}$.}
    \label{fig:Wiring diagram}
\end{figure*}

Flux-control lines, often referred to as Z-lines (associating flux with the Pauli-Z term of the Hamiltonian) consist of a series of coaxial cables, attenuators, filters, and a bias tee, as shown schematically in Fig.~\ref{fig:Wiring diagram} (a--b). While applying a direct current (DC) is straightforward, achieving the desired dynamic control is more challenging. This is due to pulse distortions, caused by impedance mismatches, and transients, introduced by capacitors and inductors. Predistortion techniques can compensate for these effects by determining the transfer function $h(t)$ of the entire flux-control line and applying its inverse $h^{-1}(t)$. However, because most of the components are located at cryogenic temperatures, the transfer function needs to be inferred indirectly using the flux-tunable element on-chip. While several such methods have been proposed~\cite{rol2020time,sung2021realization, guo2024universal,heinsoo2019digital,johnson2011controlling, baur2012realizing, butscher_master,li2024high,zhang2023characterization,jerger2019situ,zhang2025characterization,foxen2018high}, their application is complex and time-consuming, in particular as the number of qubits/couplers increases.

In our system, distortions of the flux-control pulse can be categorized as short-time and long-time effects. Short-time effects occur on the timescale of a single control-pulse operation and can arise from impedance mismatch and overshoot from inductive and capacitive elements in the flux-control line. Long-time effects are those that persist for longer than the duration of the pulse and degrade subsequent control operations; they are dominated by high-pass filtering from resistive and capacitive elements in the wiring. Figure~\ref{fig:Wiring diagram}(c) shows an example of flux-pulse deformation due to long-time transients arising from the flux control line, assuming a decay constant $\tau=10~\text{\textmu s}$. The impact of these transients on the frequency of a tunable qubit/coupler, with $\omega_{max}/2\pi=5~\mathrm{GHz}$, is illustrated in Fig.~\ref{fig:Wiring diagram}(d) emphasizing how long-time transients affect subsequent control operations. In order to reduce long-timescale effects of magnetic-flux signals, one way is to employ net-zero (NZ) pulses, defined as bipolar pulses with zero net area and a vanishing DC component~\cite{rol2019fast}. However, these effects are not completely eliminated, and, while digital filters are employed to correct these distortions~\cite{oppenheim1997signals, guo2024universal}, no simple analytical model exists, to our knowledge, that provides a systematic criterion for further suppressing the residual transients.

%In our system, distortions of the flux-control pulse can be categorized as short-time and long-time effects. Short-time effects occur on the timescale of a single control-pulse operation and can arise from impedance mismatch and overshoot from inductive and capacitive elements in the flux-control line. Long-time effects are those that persist for longer than the duration of the pulse and degrade subsequent control operations; they are dominated by high-pass filtering from resistive and capacitive elements in the wiring. Figure~\ref{fig:Wiring diagram}(c) shows an example of flux-pulse deformation due to long-time transients arising from the flux control line, assuming a decay constant $\tau=10~\text{\textmu s}$. The impact of these transients on the frequency of a tunable qubit/coupler, with $\omega_{max}/2\pi=5~\mathrm{GHz}$, is illustrated in Fig.~\ref{fig:Wiring diagram}(d) emphasizing how long-time transients affect subsequent control operations. While digital filters are typically employed to correct these distortions\cite{oppenheim1997signals, guo2024universal}, we are not aware of any simple analytical model of flux-control wiring to guide the design of flux-pulses. Typically, magnetic-flux signals are implemented by means of net-zero (NZ) pulses---defined as bipolar pulses with zero net area, resulting in a vanishing DC component\cite{rol2019fast}---because this causes the pulse response to exhibit reduced long-timescale transient effects. However, these effects are not completely eliminated.

In this paper, we model the flux-control line as a first-order series RC circuit, capturing long-time distortion. From this simple model, we derive a class of flux pulses that are resilient to first-order RC transients. The main contribution of our work lies not in the RC model itself, but in this pulse-design criterion, which highly suppresses the dominant exponential transient without requiring precise characterization of the flux-control-line time constant. We support this with both theoretical analysis and experimental demonstration. We focus on the application of flux pulses that are non-modulated and low in frequency, which are typically used to dynamically adjust flux-tunable elements,  and we limit the scope of this work to transient mitigation within the class of net-zero pulses~\cite{rol2019fast}. In particular, we apply our findings to the control of the coupler in a two-qubit quantum processor, and we demonstrate, with minimal calibration,  that the applied flux pulse is resilient to transients. We show that the proposed transient mitigation technique is robust against mischaracterization of the effective model parameters.

The paper is organized as follows. In Section~II, we describe the measurement scheme for the characterization of flux transients. 
In Section III, we present a theoretical model that establishes the conditions for generating transient-free pulses in a flux-control line dominated by a series RC circuit. 
In Section IV, we examine the robustness of the method against mischaracterization of the transient decay constant. In Section V, we discuss the class of net-zero pulses. 
Section VI concludes the paper.

\section{Measured transients}
\begin{figure*}
    \centering
    \includegraphics[scale=0.39]{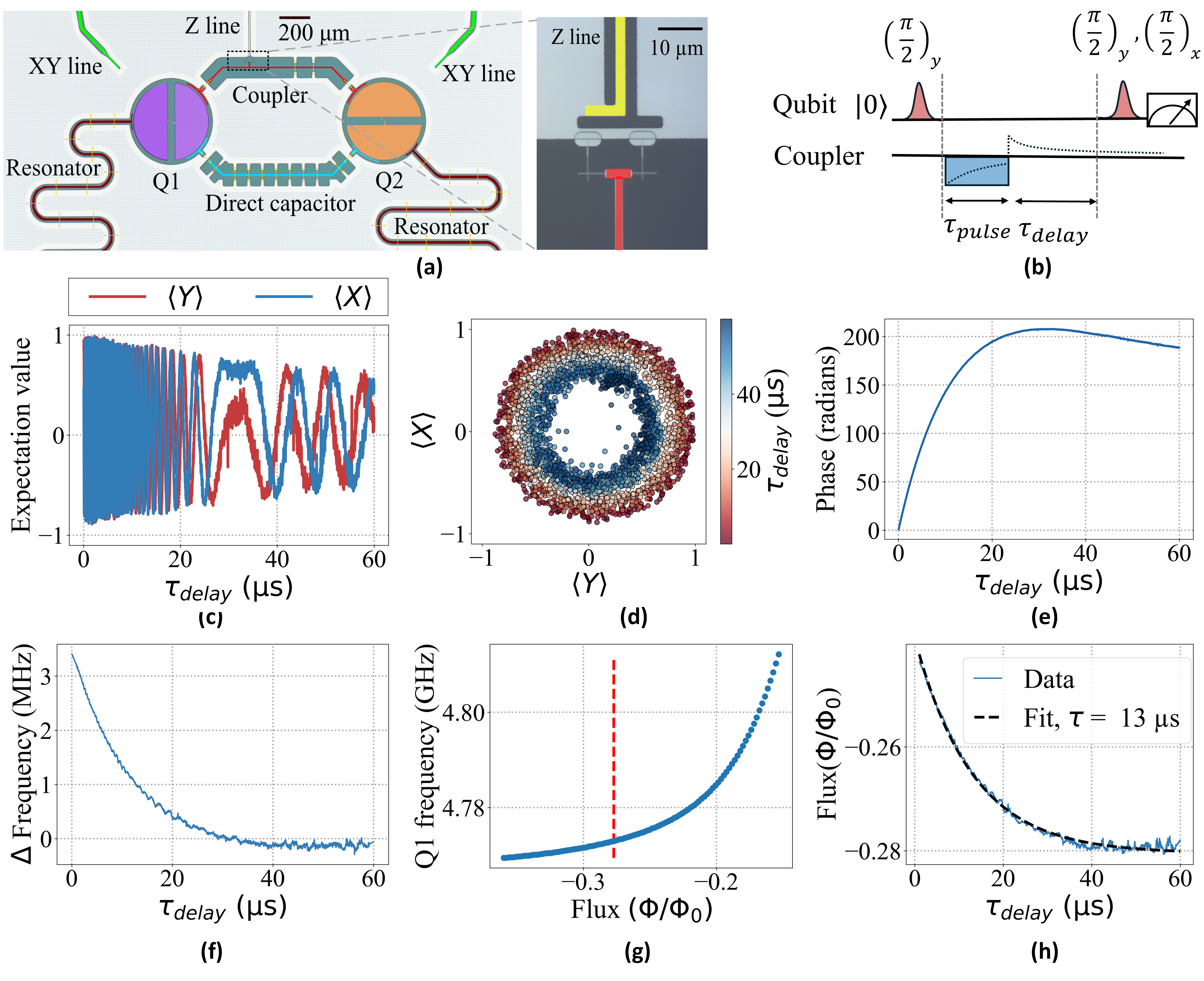}
    \caption{ Measuring flux transients. (a) Device micrograph with false coloring: two fixed-frequency qubits and one flux-tunable coupler. The right-hand side image shows a zoomed-in view of the Z-line (yellow), shorted to ground and inductively coupled to the SQUID loop of the coupler. (b) Ramsey interference pulse scheme to measure flux transients in the absence of direct readout of the coupler. We fix the coupler pulse length $\tau_{\text{pulse}}$ and vary the delay $\tau_{\text{delay}}$, and thereafter we measure the qubit's expectation values $\langle X \rangle$ and $\langle Y \rangle$. (c) Quadratures $\langle X \rangle$ (blue) and $\langle Y \rangle$ (red) vs.\@ $\tau_{\text{delay}}$ for qubit Q1: the observed oscillations indicate the presence of transients. (d) Phasor plot of the same. (e)  Extracted unwrapped phase, $\arctan(\langle Y \rangle/\langle X \rangle)$, filtered using the Savitzky-Golay method \cite{scipy_savgol_filter} to reduce high-frequency noise. (f)  Change in qubit frequency vs.\@ $\tau_{\text{delay}}$. (g) Mapping between the frequency of qubit Q1 and the normalized flux, with the idling bias shown by the red-dashed line. (h)  Extracted flux fitted with a decaying exponential function.}
    \label{fig:transients method}
\end{figure*}

Our mitigation approach is agnostic to the role of the flux-tunable element being controlled; we demonstrate it experimentally on a flux-tunable coupler, which we probe via a neighboring fixed-frequency qubit. The device is shown in Fig.~\ref{fig:transients method}(a); it consists of two fixed-frequency transmon qubits~\cite{koch2007charge} (Q1 and Q2) and a flux-tunable coupler~\cite{bengtsson2020improved,mckay2016universal}. Both qubits have dedicated readout resonators and microwave drive lines (XY), while the coupler has only a flux-control line (Z-line). This Z-line controls the flux threading the coupler's SQUID (superconducting quantum interference device), enabling tuning of the coupler's frequency.

Since the coupler has no dedicated readout resonator, we employ qubit Q1 as a probe to sense the transients in the flux control signal. When the coupler is biased near-resonantly with the qubit, any variation in the coupler's flux causes a frequency shift of the qubit. 

We use Ramsey interference of the qubit to measure the transient characteristics of the flux line~\cite{rol2020time, sung2021realization,li2024realization,glaser2024sensitivity}. Figure~\ref{fig:transients method}(b) shows the experimental procedure: two $\pi/2$ pulses are applied to qubit Q1, and a square-shaped flux pulse of duration $\tau_{\text{pulse}} =8~\text{\textmu s}$ is applied to the coupler. The qubit expectation values $\langle X \rangle$ and $\langle Y \rangle$  are measured as a function of the delay time $\tau_{\text{delay}}$. 

A system with no transients would exhibit no oscillations of $\langle X \rangle$ and $\langle Y \rangle$ for varying durations of $\tau_{\text{delay}}$, during which the flux pulse is nominally off. However, we do observe oscillations in the expectation values, as shown in Fig.~\ref{fig:transients method}(c). The phasor representation of the expectation values as a function of $\tau_{\text{delay}}$ is shown in Fig.~\ref{fig:transients method}(d). Taking the inverse tangent of $\langle Y \rangle /\langle X \rangle$, we extract how the unwrapped qubit phase changes with $\tau_{\text{delay}}$, see Fig.~\ref{fig:transients method}(e). The derivative of the phase is then used to extract the change in qubit frequency as a function of $\tau_{\text{delay}}$, as shown in Fig.~\ref{fig:transients method}(f). 

To translate the frequency changes into flux variations, we perform qubit spectroscopy as a function of flux (Fig.~\ref{fig:transients method}(g)). We then convert the frequency to flux and fit the transients to an exponentially decaying function, $A(-e^{-\tau_{\text{delay}}/\tau}+e^{-(\tau_{\text{delay}}+\tau_{\text{pulse}})/\tau})+B$, which is an approximation of the measured response. We find $\tau = 13~ \text{\textmu s}$, as shown in Fig.~\ref{fig:transients method}(h), which we identify with the characteristic time constant of an equivalent RC circuit (Fig.~\ref{fig:Wiring diagram}(b)), further discussed in Section~\ref{sec:model}. We note that although an exponential decay captures most of the transient behavior, there exist long-term effects (measured up to $300~\text{\textmu s}$), shown in Fig.~\ref{fig:long transients} in the Appendix B.

\section{Analytical model of transient-free pulses}\label{sec:model}

We model the flux-control line as a first-order RC circuit as shown in Fig.~\ref{fig:Wiring diagram}(b), neglecting inductances. This is a valid assumption for pulses that are non-modulated and low-frequency, often referred to as DC flux pulses. Although pulses may feature sharp edges, the $50\,\Omega$ cryogenic attenuators combined with the large bias tee capacitance result in a strongly overdamped regime for the flux-control line. This yields two widely separated poles: a fast inductive pole ($\tau_{\text{fast}} = L/R \approx 40\,\text{ps}$) for $L \sim 2\text{ nH}$ from the coplanar waveguide~\cite{li2025flip} and wire bonds~\cite{wenner2011wirebond} and a slow capacitive pole ($\tau_{\text{slow}} = RC \approx 10\,\mu\text{s}$) from the bias tee. Because $\tau_{\text{fast}}$ is orders of magnitude faster than pulse rise times, inductive transient decays on a negligible timescale compared with the pulse dynamics. This allows us to safely neglect $L$ and model the system as an effective RC circuit dominated by the bias tee.

In our model, given an input signal $V_{\text{in}}(t)$ and a voltage drop $V_{\text{c}}(t)$ across the capacitor, we can write 
\begin{equation}
  \tau \frac{dV_{\text{c}} (t)}{dt} + V_{\text{c}} (t) = V_{\text{in}}(t),
\label{eq:kirchoff}
\end{equation}
where $\tau = RC $. Before the application of the pulse, the circuit is assumed to be in a stationary state, such that the initial capacitor voltage is $V_c(0)=0$. Taking the Laplace transform of Eq.~\eqref{eq:kirchoff} yields
\begin{equation}
\tau sV_{\text{c}}(s)+ V_{\text{c}}(s) = V_{\text{in}}(s).
\label{eq:kirchoff_laplace}
\end{equation}

The capacitor response in the Laplace domain can be written as
\begin{equation}
V_{\text{c}}(s)
= H(s)V_{\text{in}}(s),
\label{eq}
\end{equation}
where
\begin{equation}
H(s)
= \frac{1}{\tau}\frac{1}{s+1/\tau}
\end{equation}
is the transfer function of the circuit. The corresponding impulse response is obtained by inverse Laplace transformation,
\begin{equation}
h(t)=\frac{1}{\tau}e^{-t/\tau}u(t).
\end{equation}

Consequently, the time-domain response of the capacitor voltage is given by the convolution of the input signal with the impulse response:
\begin{equation}
V_{\text{c}}(t)
= \frac{1}{\tau}e^{-t/\tau}
\int_0^t e^{t'/\tau}V_{\text{in}}(t') dt'.
\end{equation}

Assuming an input signal $V_{\text{in}}(t)$ of duration  $\tau_{\text{pulse}}$, one can write 
\begin{equation}
V_{\text{c}}(\tau_{\text{pulse}})
= e^{-\tau_{\text{pulse}}/\tau}
\left(
\frac{k}{\tau}
\right),
\end{equation}
where $k$ is defined as
\begin{equation}
\boxed{
k
= \int_0^{\tau_{\text{pulse}}}
e^{t'/\tau}V_{\text{in}}(t') dt'}.
\label{eq:kexp_Laplace}
\end{equation}

To ensure that the capacitor voltage vanishes at $t=\tau_{\text{pulse}}$, thereby eliminating the residual transient associated with the RC response, the condition $k=0$ must be satisfied.

We now give a practical example of transient-free pulse shapes, using the ansatz that $V_{\text{in}}(t)$ has a general sine-cosine Fourier series form,
 \begin{equation}
   V_{\text{in}} (t) = a_{0} +\sum_{n=1}^{N} \left[ a_{n} \cos (n \omega t) + b_{n} \sin (n \omega t )\right] ,\
   \label{eq:vin_input}
\end{equation}
with $\omega = 2 \pi /T$ and $T=\tau_{\text{pulse}}$. Using Eq.~\eqref{eq:kexp_Laplace} and imposing the condition $k=0$  give
\begin{equation}
    k = \left[ a_0 + \sum_{n=1}^{N} \frac{a_n - b_n n\omega\tau}{1 + (n\omega\tau)^2} \right] \tau \left( e^{\tau_{\text{pulse}}/\tau} - 1 \right).
\label{eq: condition2}
\end{equation}
Equation~(\ref{eq: condition2}) shows that transient cancellation is achieved when the prefactor in square brackets is equal to $0$. We refer to this prefactor as $k_\text{exp}$:
\begin{equation}
    k_{\text{exp}} = a_0 + \sum_{n=1}^{N} \frac{a_n - b_n n\omega\tau}{1 + (n\omega\tau)^2} .
\label{eq: k_exp_condition}
\end{equation}
For completeness, the generalization to an $M$-pole RC ladder is derived in Appendix~\ref{app:multipole}.

For practical pulse-shape design, one can additionally require the pulse $V_{\text{in}}(t)$ to start from zero without a discontinuous jump,
\begin{equation}
    V_{\text{in}}(0)= \sum_{n=0}^N a_n=0 \,;
\end{equation}
although this boundary condition is not required to satisfy the transient-cancellation condition. In the remainder of this work, we further impose $a_0=0$, such that the resulting waveforms have zero-time average and are part of the net-zero family of pulses.

Since we focus on the $\omega \tau > 1$ regime in this work, it is important to note that the leading order of each term in $k_{\text{exp}}$ is $(\omega \tau)^{-2}$ for the cosine modes and $(\omega \tau)^{-1}$ for the sine modes. This means that we expect a suppression of the transients independently of the particular choice of input voltage, if $\omega \tau$ is sufficiently large. We will revisit this concept in Section \ref{sec:robustness} when we discuss the robustness of our model against RC-constant mischaracterization.  

\section{Robustness against RC-constant mischaracterization}\label{sec:robustness}
Assuming one knows $\tau$ and chooses a $V_\text{in}$ that satisfies $k_\text{exp}=0$, a transient-free control pulse can be achieved. Here, we analyze $k_{\text{exp}}$ and determine whether an accurate characterization of $\tau$ is needed. Focusing on simple example cases, we showcase our theoretical findings and compare them with our experimental results. Some details of the derivations have not been included in the main text and can be found in Appendix~\ref{appendix:under_over_estimation}.

\subsection{Theoretical analysis}
To study the robustness of the method when the time constant $\tau$ is mischaracterized, we solve $k_{\text{exp}}=0$ by substituting $\tau$ with $r\, \tau$, where $r$ is an estimation factor ($r \in \mathbb{R}^+$). One possible strategy to design $V_{\text{in}}$ is to use the $N^{th}$ coefficients to cancel the contribution of the other $N-1$ harmonics contributing to $k_{\text{exp}}$. Under this change of parameters and separation of coefficients, $k_{\text{exp}}=0$ gives 
\begin{equation}
     \sum_{n=1}^{N-1} \frac{a_{n} - n \omega r \tau b_{n}}{1 + (n \omega r\tau)^{2}} +\frac{a_N^r - N \omega r \tau b_N^r}{1 + (N \omega r\tau)^{2}} = 0. \label{eq:robust1}
\end{equation}
By solving Eq.~\eqref{eq:robust1}, we obtain $a_N^r$ and $b_N^r$, where the superindexes signal that they are computed with an estimation factor $r$. Since $\tau$ is mischaracterized, these values would be incorrect for perfect cancellation of $k_{\text{exp}}$. The non-zero $k_{\text{exp}}$ becomes, 
\begin{equation}
    k_{\text{exp}}= \sum_{n=1}^{N-1} \frac{a_{n} - n \omega  \tau b_{n}}{1 + (n \omega \tau)^{2}} +\frac{a_{N}^r - N \omega  \tau b_{N}^r}{1 + (N \omega \tau)^{2}}. \label{eq:robust2}
\end{equation}
To gain an analytical understanding of cases where $\tau$ is highly overestimated, we calculate $k_{\text{exp}}$ for $r\gg1$ when $V_{\text{in}}$ contains either sine-only or cosine-only terms.  
The leading order at large $r$ gives,
\begin{alignat}{5}
   k_{\text{exp}}^{\text{cos}} &\approx& &\sum_{n=1}^{N-1} \mleft[ \frac{1}{1 + (n \omega \tau)^2} - \frac{N^2}{n^2 (1 + (N \omega \tau)^2)} \mright] a_n , \\
   k_{\text{exp}}^{\text{sin}} &\approx&-&\sum_{n=1}^{N-1} \mleft[ \frac{1}{1 + (n \omega \tau)^2} - \frac{N^2}{n^2 (1 + (N \omega \tau)^2)} \mright] n \omega \tau b_n ,
\end{alignat}

for cosine-only and sine-only $V_{\text{in}}$, respectively. 

These transient coefficients, corrected using an overestimated $\tau$, scale for large $\omega\tau$ ($r \gg 1,\,\omega\tau\gg1$) with powers $(\omega\tau)^{-4}$ for the cosine modes and $(\omega\tau)^{-3}$ for the sine modes:
\begin{align}
    \left. k_{\text{exp}}^{\text{cos}} \right\rvert_{r \gg1,\,\omega \tau \gg 1} &\approx \sum_{n=1}^{N-1} \left( \frac{1}{N^2 n^2} - \frac{1}{n^4} \right) \frac{a_n}{(\omega \tau)^4} ,
      \label{eq:kexp_cos_larg_m}
\end{align}
\begin{align}
    \left. k_{\text{exp}}^{\text{sin}} \right\rvert_{r\gg 1 ,\,\omega \tau \gg 1} &\approx \sum_{n=1}^{N-1} \left( \frac{1}{n^3} - \frac{1}{N^2 n} \right) \frac{b_n}{(\omega \tau)^3}.
    \label{eq:kexp_sin_larg_m}
\end{align}

These equations constitute one of the most important insights from this work: for pulses sufficiently faster than the time scale set by the RC-constant $\tau$, transients will be highly suppressed using the condition $k_{\text{exp}}=0$ even if we overestimate $\tau$. More details are included in Appendix~\ref{appendix:under_over_estimation}. 

 % Figure below is placed here in latex to get it at desired page
\begin{figure*}
    \centering
    \includegraphics[scale=0.39]{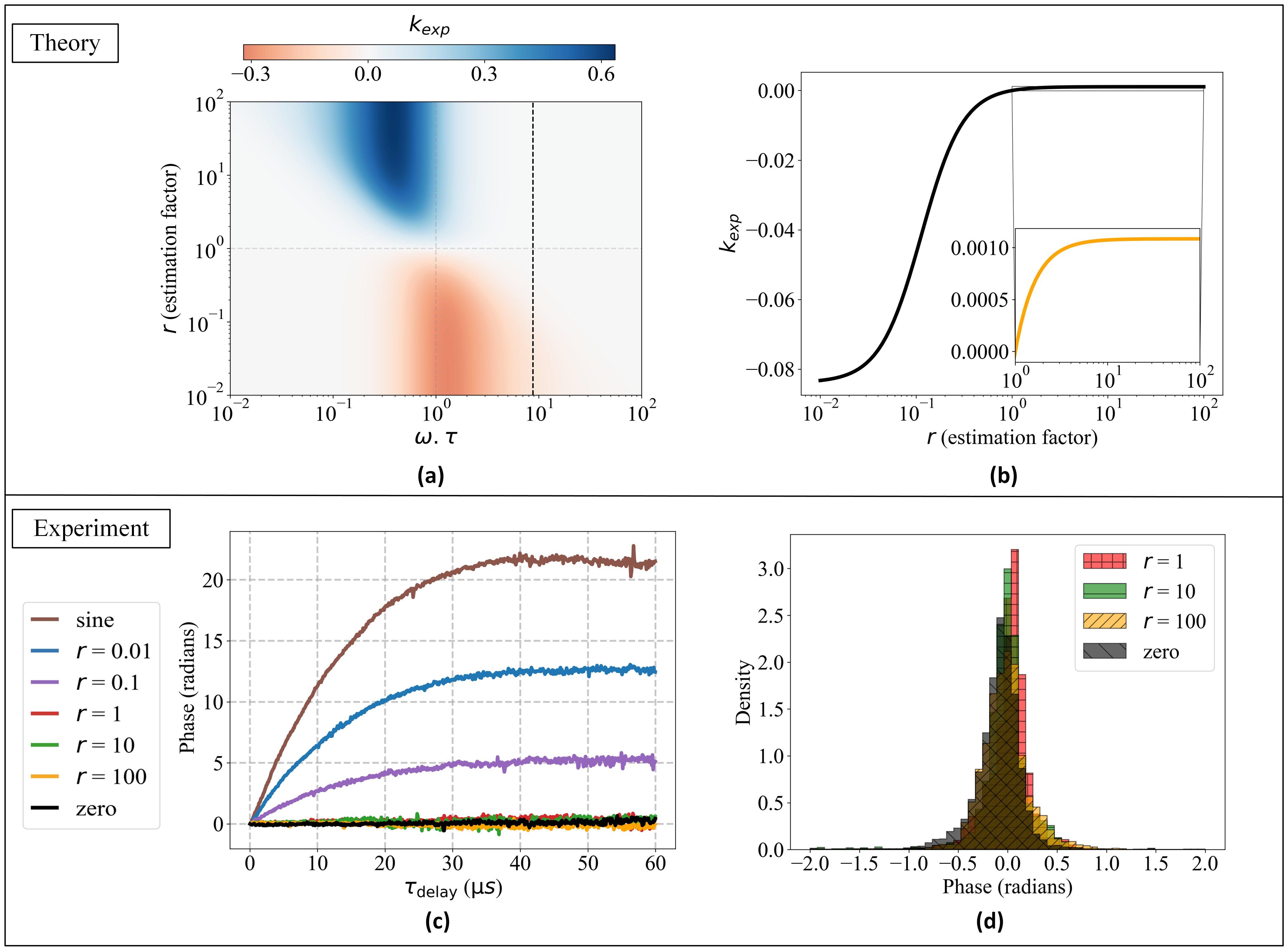}
    \caption {Robustness of the method investigated using the example case of a bi-harmonic sine signal, $V_{\text{in}}= b_1 \mathrm{sin} (\omega t)+ b_2 \mathrm{sin}(2 \omega t )$, with $\tau =11.2~\text{\textmu s}$. (a)  Analytical calculation of $k_{\text{exp}}$ as a function of $\omega \tau$ and $r$ given by Eq.~(\ref{eq:robust2}). Here $r$ is the estimation factor of $\tau$, such that the decay constant $\tau$ is mischaracterized as $r \tau$. The dashed line shows the $\omega\tau$ value used in measurements. (b) Line cut along the dashed line in (a), demonstrating that the overestimation of $\tau$ preserves the mitigation of transients. (c) Experimental verification: Phase acquired by qubit Q1 as a function of $\tau_{\text{delay}}$, extracted via the Ramsey-like method in Fig.~\ref{fig:transients method}(a) for different pulse shapes: zero (reference pulse), single-frequency pulse $V_{\text{in}}$ (sine), and bi-harmonic pulse $V_{\text{in}}$ for different values of the estimation factor $r$. The values $r=10$ (green) and $100$ (orange) indicate over-estimation while $r=0.1$ (purple) and $0.01$ (blue) represent under-estimation, and $r=1$ shows the exact value of $\tau$ used for the bi-harmonic pulse. The single-frequency pulse $V_{\text{in}}$ (brown) results in the highest acquired phase, thus representing greater transients. (d) Histograms of the acquired phase for $r=1$, 10, 100, and the reference pulse (zero). Note that the $\tau =11.2~\text{\textmu s}$ used in this experiment is slightly different from the one estimated in Fig.~\ref{fig:transients method}(h) ($\tau = 13~\text{\textmu s}$); this discrepancy is due to parameter extraction with a refined fitting method at a later stage.}
    \label{fig:robusteness}
\end{figure*}

The discussion so far has been rather abstract. We can evaluate the effectiveness of the method in a simpler and quantitative way by taking the example of a bi-harmonic sine signal, defined by the following input voltage,
\begin{equation}
V_{\text{in}}(t)= b_1\, \mathrm{sin} (\omega t)+ b_2 \, \mathrm{sin}(2 \omega t ).
\label{eq: bi-harmonic vin}
\end{equation}

If one knows the time constant $\tau$, full transient cancellation is achieved when $k_{\text{exp}}=0$ is satisfied, giving the resulting relationship between the two Fourier coefficients,
\begin{equation}
    b_2=\frac{-b_1}{2} \frac{1+(2\omega \tau)^2}{1+(\omega \tau)^2} . \label{eq:cancel b2}
\end{equation}

However, if we overestimate $\tau$ with estimation factor $r$, as discussed above, we end up with a coefficient 
\begin{equation}
    b_2^r=\frac{-b_1}{2} \frac{1+(2\omega r\tau)^2}{1+(\omega r \tau)^2} , \label{eq:cancel b2 overestimate}
\end{equation}
that is formally equivalent to $b_2$ in Eq.~\eqref{eq:cancel b2}, but with $\tau$ replaced by $r \tau$. Here again, we use the $r$ superindex to signal that the coefficient was obtained with the overestimated RC-constant.

Figure~\ref{fig:robusteness}(a) shows $k_{\text{exp}}$ as a function of $r$ and $\omega \tau$ when we set $b_1 =1$ and $b_2=b_2^r$.It is interesting to note that $k_{\text{exp}}$ is small for both large and small $\omega \tau$ despite the large variation in $r$. For $\omega \tau \approx 1$, there is instead a significant error arising from the mischaracterization of $\tau$. Figure~\ref{fig:robusteness}(b) shows the exponential coefficient $k_{\text{exp}}$ as a function of $r$ for $\omega \tau = 8.79~\mathrm{rad}$, which we use in measurements presented later. It can be seen that for large values of $r$, $k_{\text{exp}}$ is negligibly small and therefore the overestimation of $\tau$ still leads to a significant suppression of first-order RC transients. 

We can understand this suppression if we take the leading order in large $r$ of the solution to condition $k_{\text{exp}}$ (Eq.\eqref{eq:cancel b2 overestimate}), 
\begin{equation}
    b_2^{r\gg 1} \approx -2 b_1.
\end{equation}
With this, the transient coefficient $k_{\text{exp}}$ becomes
\begin{equation}
    k_{\text{exp}}=\frac{3 b_1 \omega \tau}{1+ 5\omega^2 \tau^2 +4 \omega^4 \tau^4}. \label{eq: sine case exponen}
\end{equation}
For large $\omega \tau$,  $k_{\text{exp}}$ scales as a negative cubic power, thus making the transients negligible even with a mischaracterized $r$.

It is interesting to compare the coefficient $k_{\text{exp}}$ in Eq.~(\ref{eq: sine case exponen}) to what would be obtained in the case of a single-frequency sine signal, $V_{\text{in}} = b_1 \mathrm{sin(\omega t)}$. In this case, the transient coefficient would be
\begin{equation}
    k_{\text{exp}} = \frac{-\omega \tau b_1}{1 + \omega^2 \tau^2} .
\end{equation}
Although we are comparing two different pulse shapes, this still provides insight. By looking at the powers of $\omega\tau$ alone, it is evident that, for $\omega \tau \gg 1 $, an overestimation of the time constant provides a stronger suppression of transients for a bi-harmonic sine signal in comparison to a standard single-frequency pulse shape.

\subsection{Experimental verification}
We experimentally verify the robustness of our technique using the bi-harmonic sine signal, $V_{\text{in}}$ = $b_1\, \mathrm{sin} (\omega t) + b_2^r\, \mathrm{sin} (2 \omega t)$, discussed earlier. We characterize the transients using the method outlined in Section II. The acquired phase after the flux pulse is treated as the figure of merit for the transients (the acquired phase refers to the subtracted phase offset from the extracted unwrapped phase). We measure the acquired phase as a function of $\tau_{\text{delay}}$ for a variety of pulses of length $\tau_{\text{pulse}}=8~\text{\textmu s}$: a zero-amplitude pulse for reference, a single-frequency sine pulse, and bi-harmonic sine pulses with different values of the estimation factor $r$. 
The results are shown in Fig.~\ref{fig:robusteness}(c), demonstrating that an overestimation of the time constant leads to negligible transients. The measured phases for $r=10$ and $100$ almost overlap with those for $r=1$ and the zero pulse. Note that, as predicted by the theory, underestimating $\tau$ leads to increased transients as shown for $r = 0.1$ and $0.01$ (see Fig.~\ref{fig:robusteness}(b)). Interestingly, our measurement shows that a single-frequency sine pulse results in even larger transients. The broad range of $r$ explored here serves as a stress test of the method's robustness, and is motivated by scenarios in which $\tau$ is only coarsely known, e.g., from nominal specifications rather than dedicated characterization.

We further investigate whether the phase distributions $r=1$, $10$, and $100$ are statistically distinguishable from that of the zero-amplitude pulse. For these cases, we repeat the phase measurement in Fig.~\ref{fig:robusteness}(c) 11 times. The phase values from these datasets are then combined for each case to obtain the histogram plot in Fig.~\ref{fig:robusteness}(d). The results show that overestimating the time constant used to design the bi-harmonic sine pulse does not produce a statistically significant shift in the phase distribution relative to the zero-amplitude case. The corresponding mean values and standard deviations are reported in Appendix \ref{app:phase_histogram}

%TO bring it on correct page
\begin{figure*}
    \centering
    \includegraphics[scale=0.39 ]{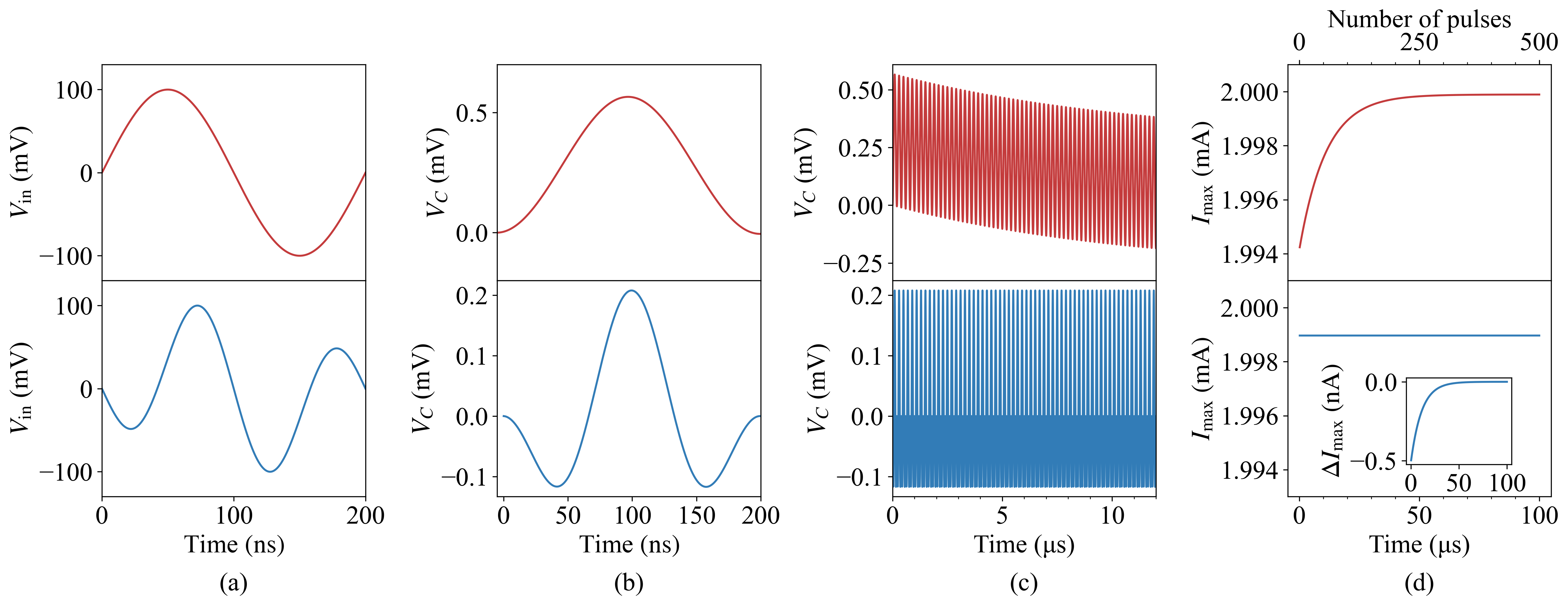}
    \caption{ Comparison between using either a single-frequency sine signal (red curve) or a bi-harmonic sine signal (blue curve), highly overestimating the time constant with $b_2 = -2~ b_1$. We consider that the system has an RC time constant $\tau= 11.2~\text{\textmu s}$. (a) Input pulse shape $V_{\text{in}}$ for one period. (b) Capacitor response $V_{\text{c}}(t)$ for one period. (c) Capacitor response $V_{\text{c}}(t)$ over multiple periods. The resulting $V_{\text{c}}(t)$ in the case of the single-frequency sine signal has transients, while in the case of the bi-harmonic sine signal transients are almost negligible. (d)  Maximum current per pulse $I_{\text{max}}$, assuming $R = 50~\Omega$, showing non-negligible transient for the single-frequency sine signal, while the bi-harmonic signal remains flat (inset shows sub-nA residual variations $\Delta I_{\text{max}.})$}
    \label{fig:robusteness1}
\end{figure*}
\section{Discussion of net-zero pulses}

A strategy to mitigate long-time transients is to employ NZ pulses. We have shown that a sub-class of NZ pulses can provide even stronger suppression. In our experiment, we tested input signals having the form of single-frequency and bi-harmonic sine pulses (see Fig.~\ref{fig:robusteness}(c));  
while both are net-zero considering $V_{\text{in}}$, the amplitude at the generator (Fig.~\ref{fig:robusteness1}(a)), we show that they produce different transients in the output flux that reaches the device under test. 

To understand this discrepancy, we consider the circuit in Fig.~\ref{fig:Wiring diagram}(b) and compare the voltage drop $V_{\text{c}}(t)$ across the capacitor (see Fig.~\ref{fig:robusteness1}(b)) \cite{footnote1}. 
For the bi-harmonic sine pulse, $V_{\text{c}}(t)$ is (approximately) net-zero, while for the single-frequency sine pulse it is not. 
In Fig.~\ref{fig:robusteness1}(c), we compare the capacitor's response over multiple periods, showing that $V_{\text{c}}(t)$ for the single-frequency signal has significant transients in comparison to that of the bi-harmonic signal. 
This elucidates that a net-zero input pulse is not a sufficient condition to minimize RC transients: an effective suppression of transients also requires the capacitor's response $V_{\text{c}}(t)$ to be net-zero. As the resulting current is responsible for inducing flux in the circuit, we also show the transients affecting the current by plotting the maximum current ($I_\text{max}$) per pulse period in Fig.~\ref{fig:robusteness1}(d). 

 \section{Conclusion}
In this work, we introduce a pulse-design criterion, derived from a simple first-order RC model of the flux-control line, that highly suppresses long-time transients without requiring precise knowledge of the line's time constant. We validate its effectiveness both theoretically and experimentally on a flux-tunable coupler, demonstrating its robustness against mischaracterization of the model parameters. This robustness enables efficient experiment tune-up without detailed transient characterization. Rather than replacing predistortion-based correction, our approach offers a complementary and calibration-free strategy for transient mitigation, becoming advantageous as the number of flux-control lines increases with processor size. We believe that the proposed method provides a practical solution for improving flux control in superconducting quantum processors.

\begin{acknowledgments}
We would like to thank Simon Pettersson Fors, Christopher W. Warren, Tomas McKelvey, Simone Gasparinetti, and Per Delsing for useful discussions. This research was financially supported by the Knut and Alice Wallenberg Foundation through the Wallenberg Center for Quantum Technology (WACQT) and by the EU Flagship on Quantum Technology HORIZON-CL4-2022-QUANTUM-01-SGA project 101113946 OpenSuperQPlus100. Device fabrication was performed at Myfab Chalmers.
\end{acknowledgments}

\section*{Data Availability}
All relevant data and figures supporting the main conclusions of the document are available on Zenodo \cite{aggarwal_2025_transients_flux}. Please refer to Anuj Aggarwal at anuja@chalmers.se if needed.
\section*{Conflict of Interest}
The authors have no conflicts to disclose.

\section*{Author contributions}
A.A. designed the chip, performed the measurements and analysis of the results; A.A. and J.F.P. conceptualized the problem and developed the theory. A.A., M.R., A.N., E.H., J.B., A.F.R. participated in the fabrication of the device. T.A, D.S, H.J., R.R, and M.F.G helped in analysis of the data. A.A., J.F.P. wrote and J.By, G.T. edited the manuscript. J.By., and G.T. provided supervision and guidance during the project. All authors contributed to the discussions and interpretations of the results.
\appendix

\section{Measurement setup}
\begin{figure}[htbp]
    \centering
    \includegraphics[scale=0.51]{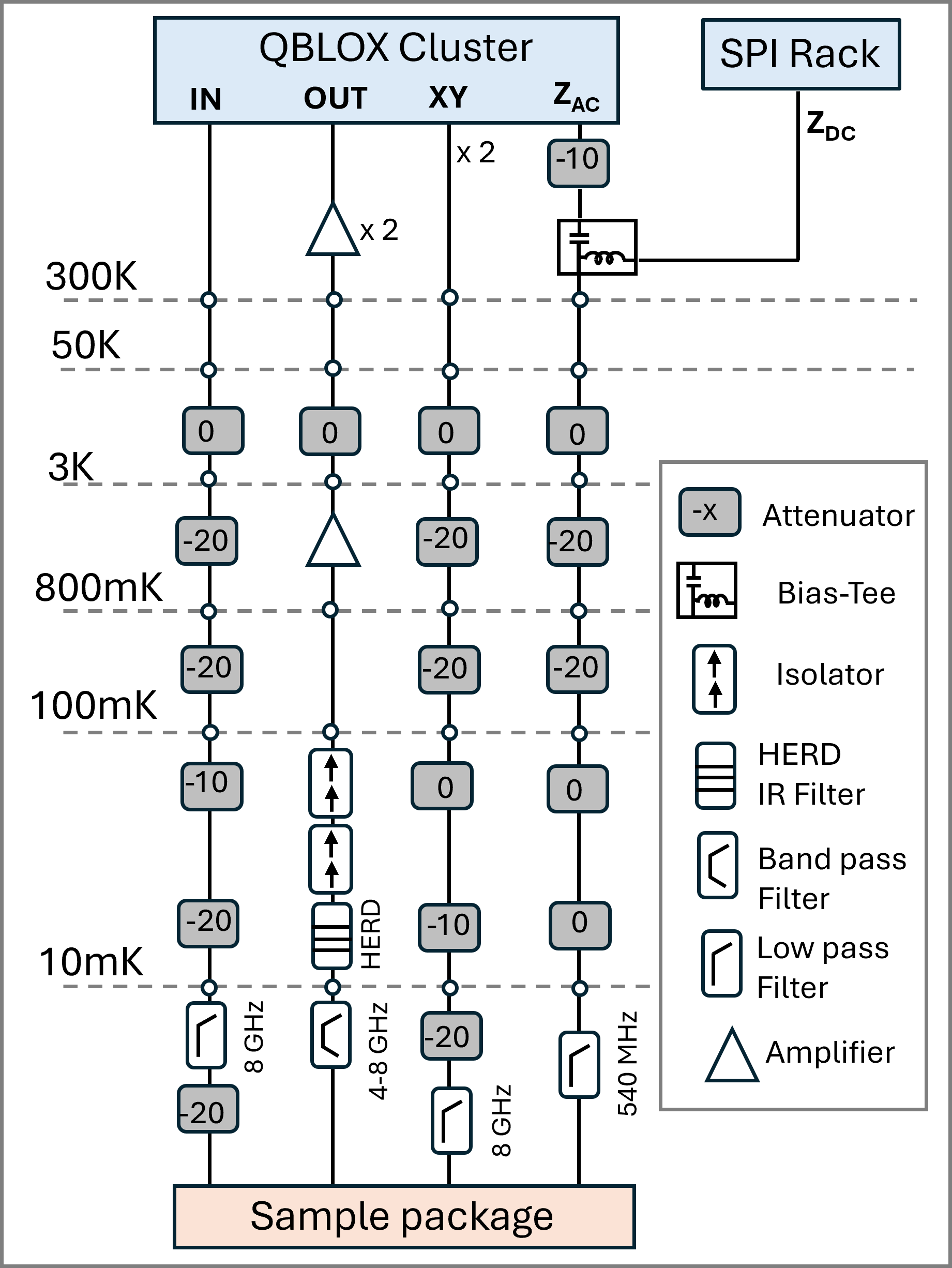}
    \caption{Schematic diagram of the measurement setup. The control lines are defined as follows: XY - qubit microwave drive line, IN - readout input line, and OUT - readout output line. $Z_{\text{AC}}$ (dynamic) and $Z_{\text{DC}}$ (static) flux control lines are coupled via a bias-tee at room temperature. Attenuator values are specified in dB units. HERD is an IR blocking filter, see~\cite{rehammar2023low}.}
    \label{fig:wiring}
\end{figure}

The schematic diagram of the measurement setup is shown in Fig.~\ref{fig:wiring}. The device is mounted at the 10-mK stage of a Bluefors LD250 dilution refrigerator. The quantum processor consists of two floating transmon qubits coupled via a coupler (frequency-tunable transmon). Each qubit is coupled to individual readout (quarter-wavelength) resonators and microwave drive lines. The coupler's SQUID is inductively coupled to a flux control line. The coupler does not have its own dedicated readout resonator. In the experiments, we use qubit Q1 to measure flux-pulse transients. The device parameters of qubit Q1 and coupler are listed in Table~\ref{table:device_param}. 

\begin{table}[htbp]
\caption{Qubit Q1 parameters and coherence properties at a coupler flux bias $\Phi=-0.278~\mathrm{\Phi_0}$.} 
\label{table:device_param}
\begin{ruledtabular}
\renewcommand{\arraystretch}{1.3} 
\begin{tabular}{llc}
\multicolumn{2}{l}{\textbf{Parameter}} & \textbf{Q1} \\ 
\hline
Qubit frequency & $\omega_q/2\pi\, (\mathrm{GHz})$ & 4.7730 \\
Qubit anharmonicity & $\alpha/2\pi\, (\mathrm{MHz})$ & $-240$ \\
Resonator frequency & $\omega_r/2\pi \,(\mathrm{GHz})$ & 6.9975 \\
Resonator linewidth & $\kappa_r/2\pi\,(\mathrm{kHz})$ & 409 \\
Relaxation time & $\overline{T}_1 \,(\mu\mathrm{s})$ & 78 \\
Decoherence time & $\overline{T}^*_2 \,(\mu\mathrm{s})$ & 75 \\
\hline
Coupler freq.\@ $(\Phi=0)$ & $\omega_c /2\pi\, (\mathrm{GHz})$ & 4.8575 \\
Qubit-coupler coupling & $g/2\pi\, (\mathrm{MHz})$ & 77 \\
\end{tabular}
\end{ruledtabular}
\end{table}

The device is fabricated on a highly resistive silicon substrate. After chemical cleaning, an aluminum ground layer is deposited, and the device pattern, excluding Josephson junctions and air bridges, is defined via optical lithography, followed by etching to form resonators, the feedline, control lines, and qubit capacitance. Josephson junctions are then patterned using e-beam lithography and fabricated through double-angle shadow evaporation of aluminum with in situ oxidation. After lift-off, lithography and deposition of a patch establish galvanic connections between the Josephson junction electrodes and the wiring layer. Finally, air bridges are added over coplanar-waveguide (CPW) structures using optical lithography, aluminum deposition, and etching to suppress slot-line modes. More details can be found in~\cite{biznarova2024mitigation,chayanun2024characterization}.

We use a Qblox cluster as the primary control hardware. It consists of QCM-RF, QRM-RF, and QCM modules. QCM-RF controls the single-qubit gates via the XY drive line of the qubit. QRM-RF is a multiplexed readout module that has two ports, one for sending the readout pulse and another to digitize the received readout signal. It is connected to the IN and OUT lines of the readout. QCM is a control module that can send arbitrary pulses from DC to 400 MHz. It is used on the $Z_{\text{AC}}$ line to send flux pulses to the coupler. We also use a Qblox SPI rack that consists of S4g current source modules. It is connected to $Z_{\text{DC}}$ and is used to set the static magnetic flux bias on the coupler. The $Z_{\text{DC}}$ and $Z_{\text{AC}}$ controls of the flux are coupled together with a bias tee (Marki- BT0025) at room temperature. Measurements were partially performed utilizing the software suite 'SuperConducting Qubit Tools (SCQT)' from Orange Quantum Systems.

\section{Transient measurements up to 300~{\textmu s}}
\begin{figure}
    \centering
    \includegraphics[scale=0.3]{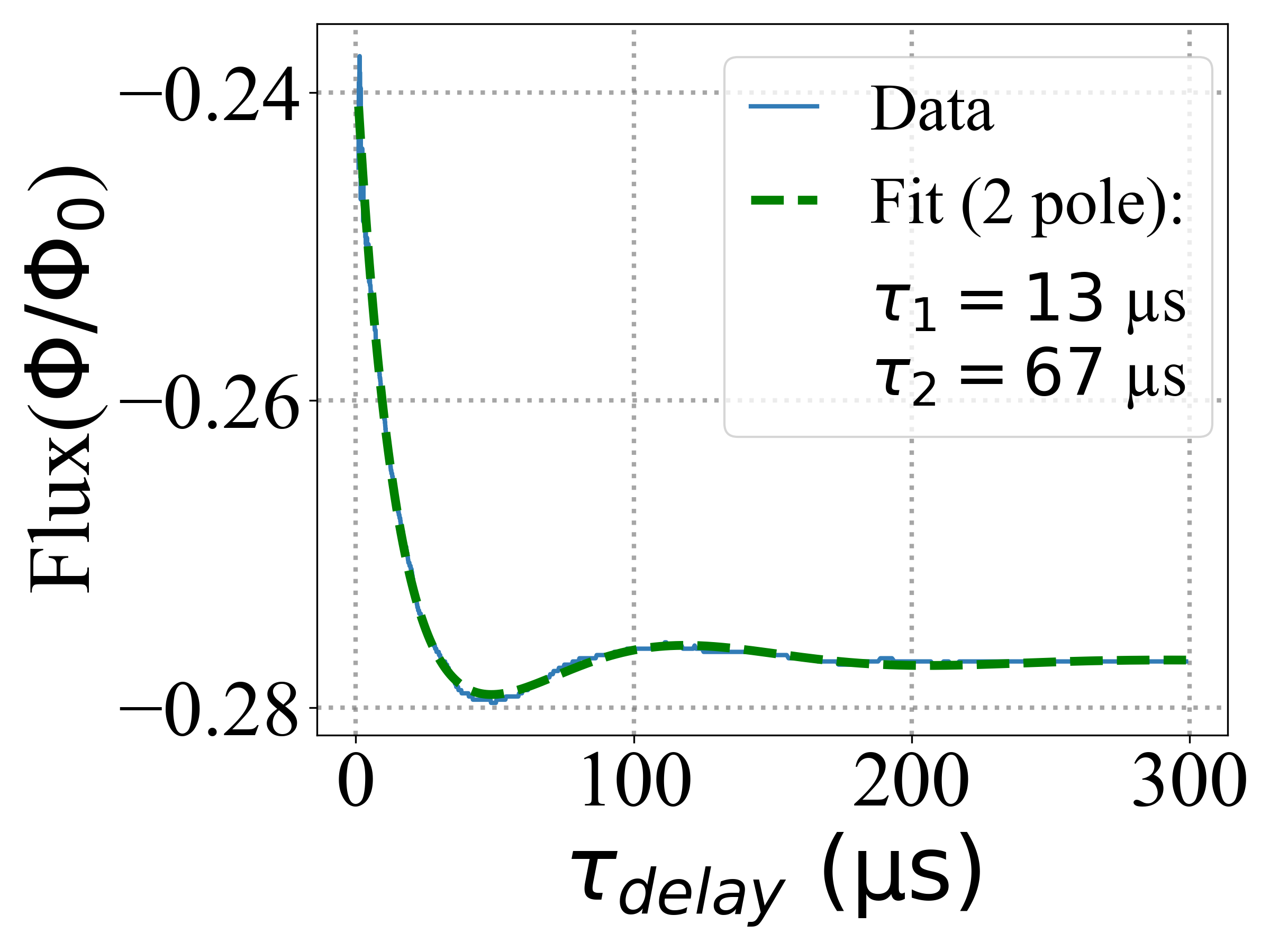}
    \caption{Extracted flux transients measured with a square pulse of $8~ \text{\textmu s} $ duration as a function of a $\tau_{\text{delay}}$ up to $300~\text{\textmu s}$ (cf.\@ Fig.~\ref{fig:transients method}(e)). The data is fitted to a sum of exponential decay and  damped oscillation, Eq.\eqref{eq:fit_2_pole}, revealing two poles with decay time  $\tau_1 = 13\,\mu\mathrm{s}$ and $\tau_2 = 67\,\mu\mathrm{s}$.}
    \label{fig:long transients}
\end{figure}

In Section II of the main text, we show the measured transients up to $60~\text{\textmu s}$, where we fit a single exponential decay to estimate the decay constant $\tau$, later related to the equivalent RC constant. Figure~\ref{fig:long transients} shows the measured transients up to $300 ~\text{\textmu s}$. Fitting this extended trace to a sum of exponential decay and  damped oscillation,
\begin{align}
y(t) =\;& -a_1\, e^{-t/\tau_1}\left(1-e^{-t_{\mathrm{pulse}}/\tau_1}\right) \nonumber \\
&- a_2\, e^{-t/\tau_2} \bigg[\cos(\omega t+\phi) \nonumber \\
&\hspace{2em} - e^{-t_{\mathrm{pulse}}/\tau_2}\cos\big(\omega(t+t_{\mathrm{pulse}})+\phi\big)\bigg] + y_{\mathrm{offset}}
,\label{eq:fit_2_pole}
\end{align}
we find a second pole corresponding to a characteristic decay time of $67 ~\text{\textmu s}$  \cite{guo2024universal,an2026zz}, likely caused by an impedance mismatch along the flux line. The first pole dominates the initial system response: its modal amplitude ($a_1 = -0.0856$) accounts for roughly $79\%$ of the total transient magnitude, compared with $21\%$ for the second pole ($a_2 = 0.0223$). Furthermore, the first pole decays roughly five times faster, with $\tau_1 = 13\,\mu\mathrm{s}$ compared to\ $\tau_2 = 67\,\mu\mathrm{s}$. Since these results indicate that the single-exponential model captures the dominant transient response, we use the corresponding first-order model as an effective description of the dominant transient, while recognizing that additional long-time transients are not captured by this simple model. Alternative protocols based on pulsed spectroscopy, not limited by qubit
coherence, should be exploited for measuring transients at longer time
scales~\cite{johnson2011controlling,barends2014superconducting,hellings2025calibrating}.

\section{Coherence data of Qubit 1}
\begin{figure}
    \centering
    \includegraphics[scale=0.51]{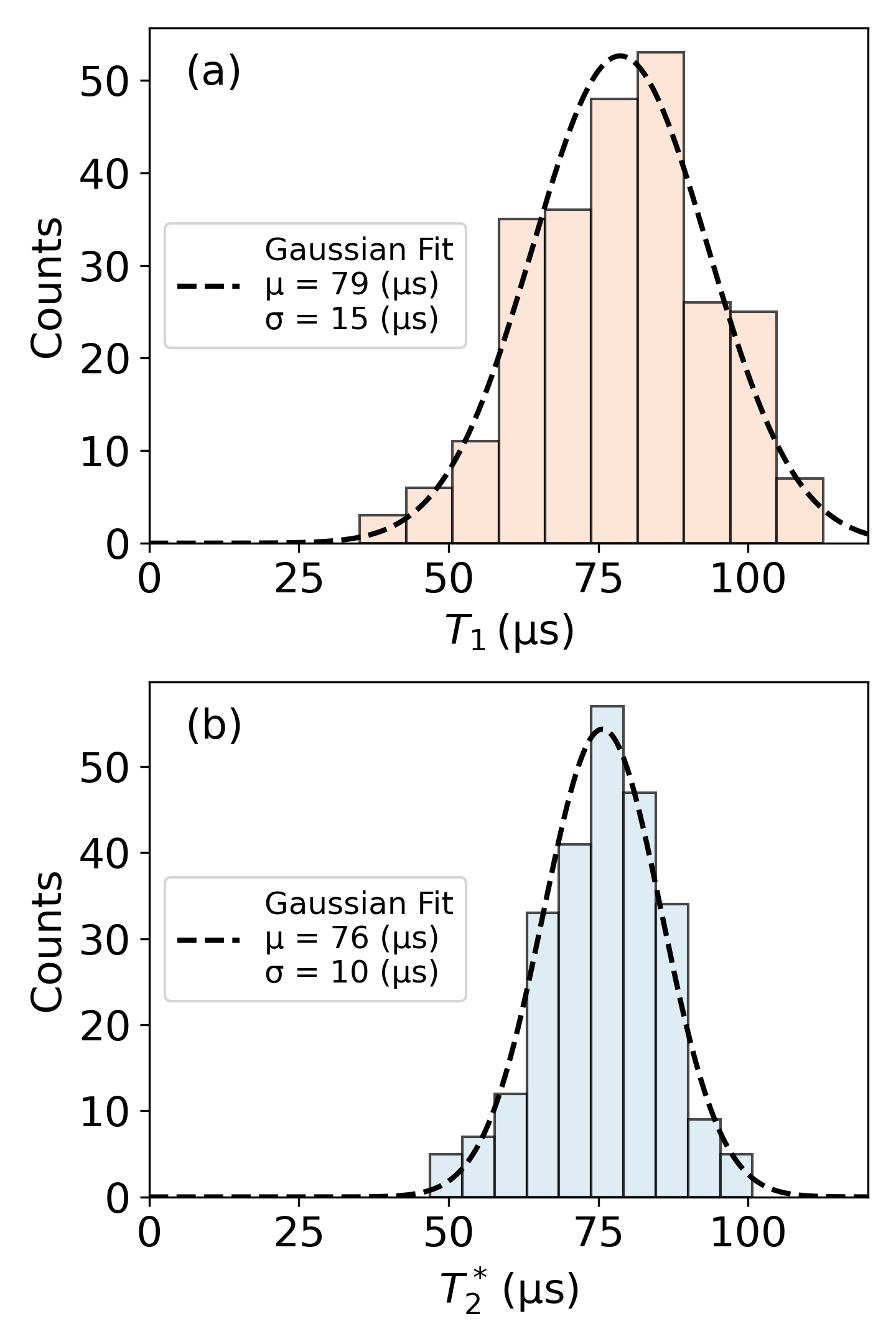}
    \caption{ Histogram distribution of (a) the $T_1$ relaxation time  and (b) the $T^*_2$ free-induction decay  (Ramsey) decoherence time of  qubit Q1, measured over 8 hours.}
    \label{fig: coherence hist}
\end{figure}

To statistically analyze the relaxation time ($T_1$) and Ramsey decoherence time ($T^*_2$) of qubit Q1, 250 measurements of each were taken over eight hours. The resulting histograms are shown in Fig.~\ref{fig: coherence hist}. 

\section{Qubit 1 spectroscopy as a function of flux bias on coupler}
In this device, qubit Q1 is coupled to a flux-tunable coupler as shown in Fig.~\ref{fig:transients method}(a). The coupler’s transition frequency is modulated by the magnetic flux threading its SQUID and is given by
\begin{equation}
    \omega_\text{c}= \omega_{\text{max}} \sqrt{|\cos (\pi \Phi/\Phi_0)|},
\end{equation}
where $\omega_{\text{max}}$ is the coupler's frequency at zero effective bias, and $\Phi_0$ is the magnetic flux quantum. 

\begin{figure}
    \centering
    \includegraphics[scale=0.35]{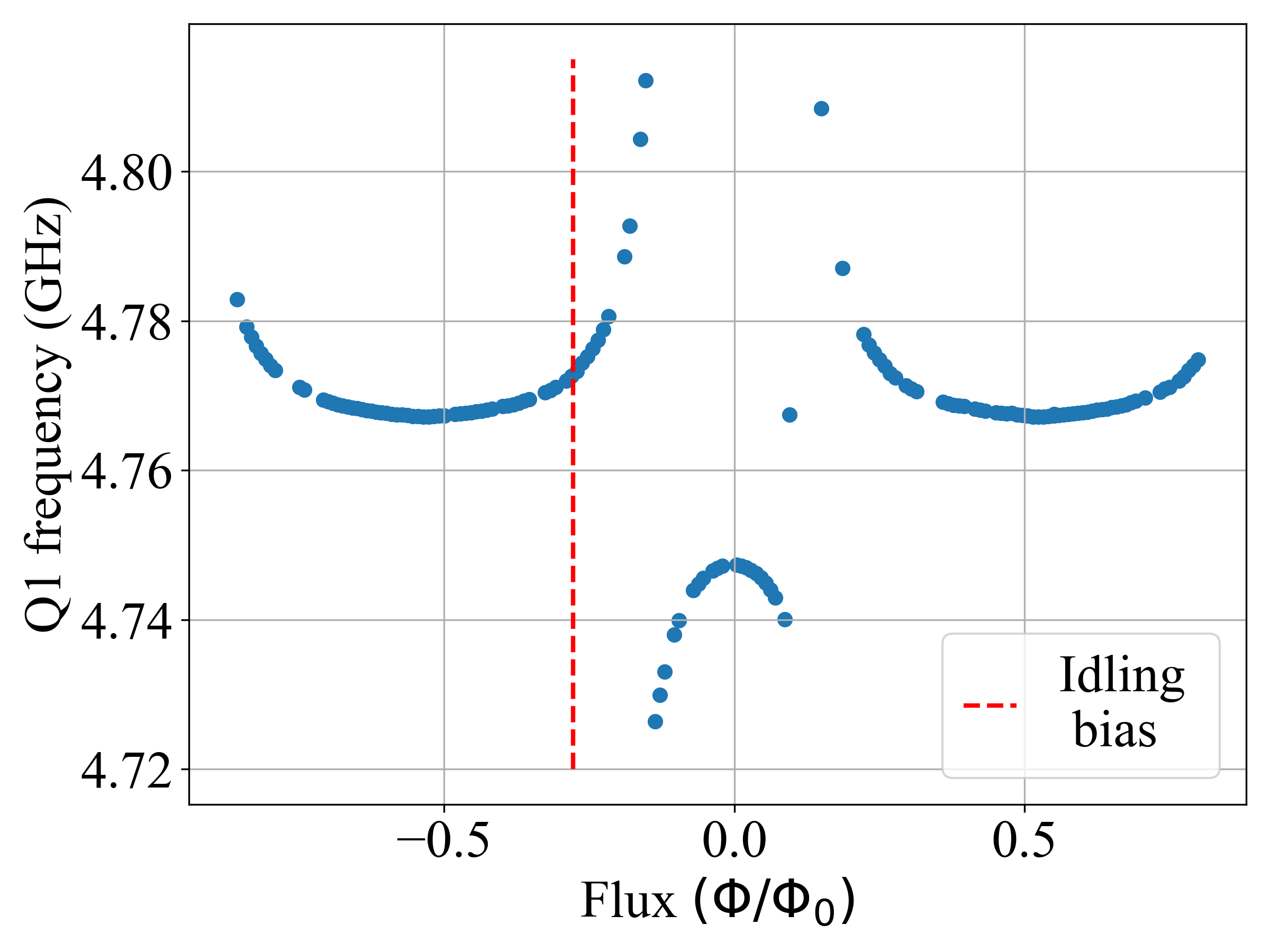}
    \caption{Frequency of qubit Q1 as a function of the coupler's flux bias. The red-dashed line shows the idling coupler bias used in the measurements.}
    \label{fig: coupler spec}
\end{figure}

Tuning the coupler’s frequency results in a change of the frequency of Q1 due to the interaction between the qubit and coupler. Figure~\ref{fig: coupler spec} shows the frequency of Q1 as a function of the coupler flux $\Phi$. When the coupler is far from resonance with the qubit ($\Phi \approx 0.5\, \Phi_0$), the frequency of Q1 remains nearly constant. However, as the coupler approaches the transition frequency of Q1, an avoided crossing occurs, causing a significant frequency shift. This avoided-crossing region is leveraged in transient measurements to enhance sensitivity. The red-dashed line in the plot marks the coupler’s bias point during idling.

\section{Single-shot readout fidelity}
\begin{figure}
    \centering
    \includegraphics[scale=0.54]{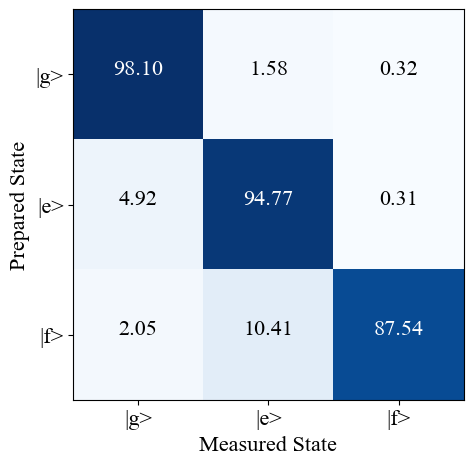}
    \caption{Readout confusion matrix for Q1. The readout fidelity is about $93\%$.}
    \label{fig: ssro}
\end{figure}

Transient characterization was performed using single-shot measurements, with state discrimination between the ground, first excited, and second excited states,  $|g \rangle, \, |e \rangle, \, |f \rangle$. The measured readout confusion matrix for qubit Q1 is shown in  Fig.~\ref{fig: ssro}. The readout fidelity is approximately $93\%$.

\section{Pulse reconstruction}
\begin{figure}[htbp]
    \centering
    \includegraphics[scale=0.34]{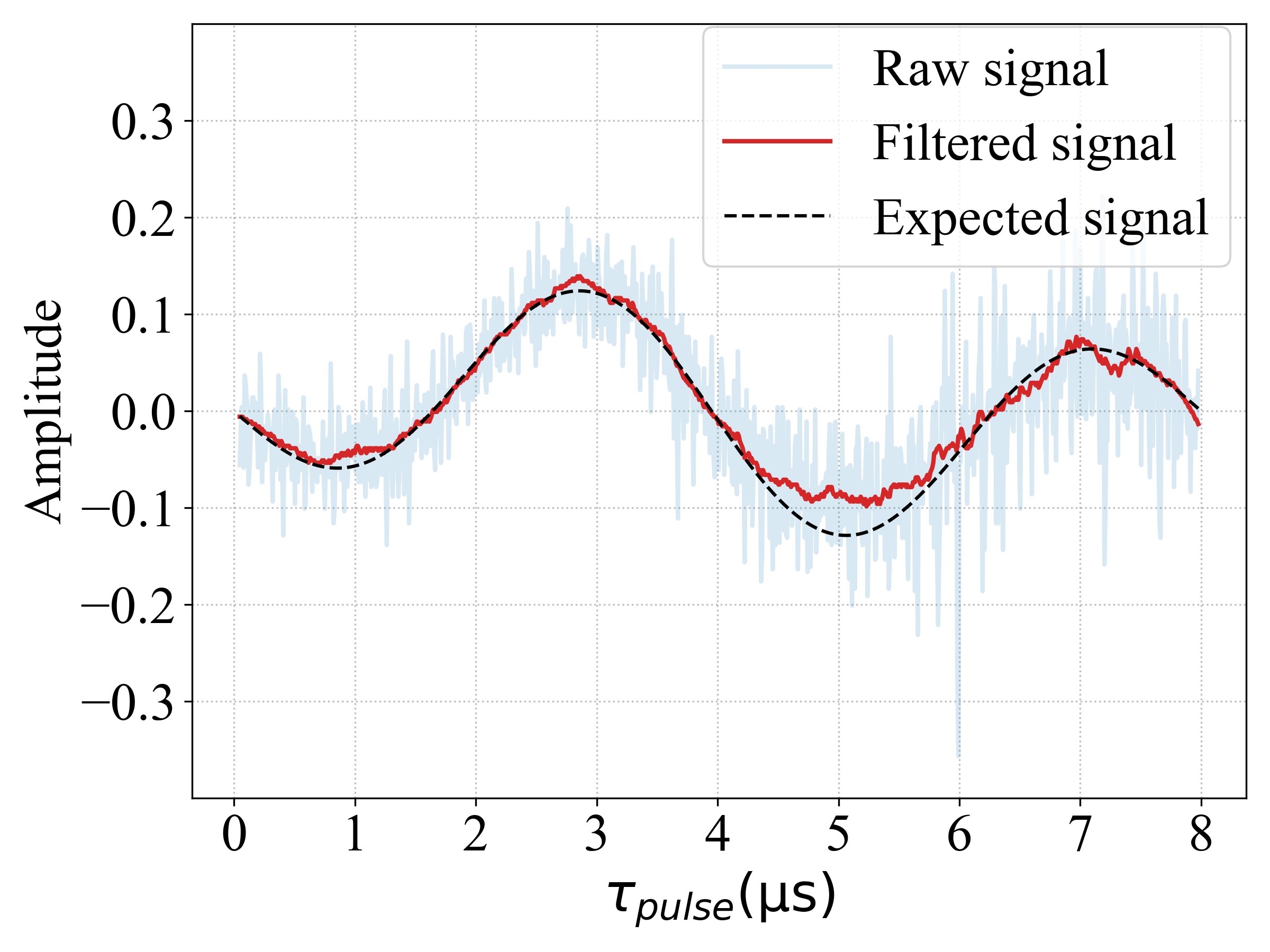}
    \caption{Pulse reconstruction for a bi-harmonic sine pulse with $r \,= \,100$, $\tau \, =11.2~\textrm{\textmu s}$, $b_1\, =\,1$, and $b_2$ given by Eq.~(\ref{eq:cancel b2 overestimate}). The reconstruction is performed with the cryoscope technique \cite{rol2020time}.}
    \label{fig: pulse reconstructiom}
\end{figure}

To verify that the desired pulse shape is reaching the coupler on the chip, we used the cryoscope technique~\cite{rol2020time}. We test the bi-harmonic flux pulse from Eq.~(\ref{eq: bi-harmonic vin}), with $b_1\,=\,1$, $b_2$ extracted from Eq.~(\ref{eq:cancel b2 overestimate}) by setting $r\,=\,100$, and $\tau = 11.2~\text{\textmu s}$. Figure~\ref{fig: pulse reconstructiom} shows that the expected and measured pulse shapes are in good agreement, except for some regions on the negative side. This occurs because, for a bipolar pulse, accurately capturing both positive and negative amplitudes requires the signal-to-noise ratio (SNR) to be comparable on both sides. In our case, the SNR is higher on the positive side and lower on the negative side. As a result, the negative side is smoothed out during averaging and filtering. One can optimize the idling bias to obtain a similar SNR on both sides. In these measurements, we choose the idling coupler bias  $\Phi=-0.227 \, \mathrm{\Phi_0}$.

 \section{Two qubit adiabatic CZ gate using a bi-harmonic pulse}

\begin{figure}[htbp]
    \centering
    \includegraphics[scale=0.34]{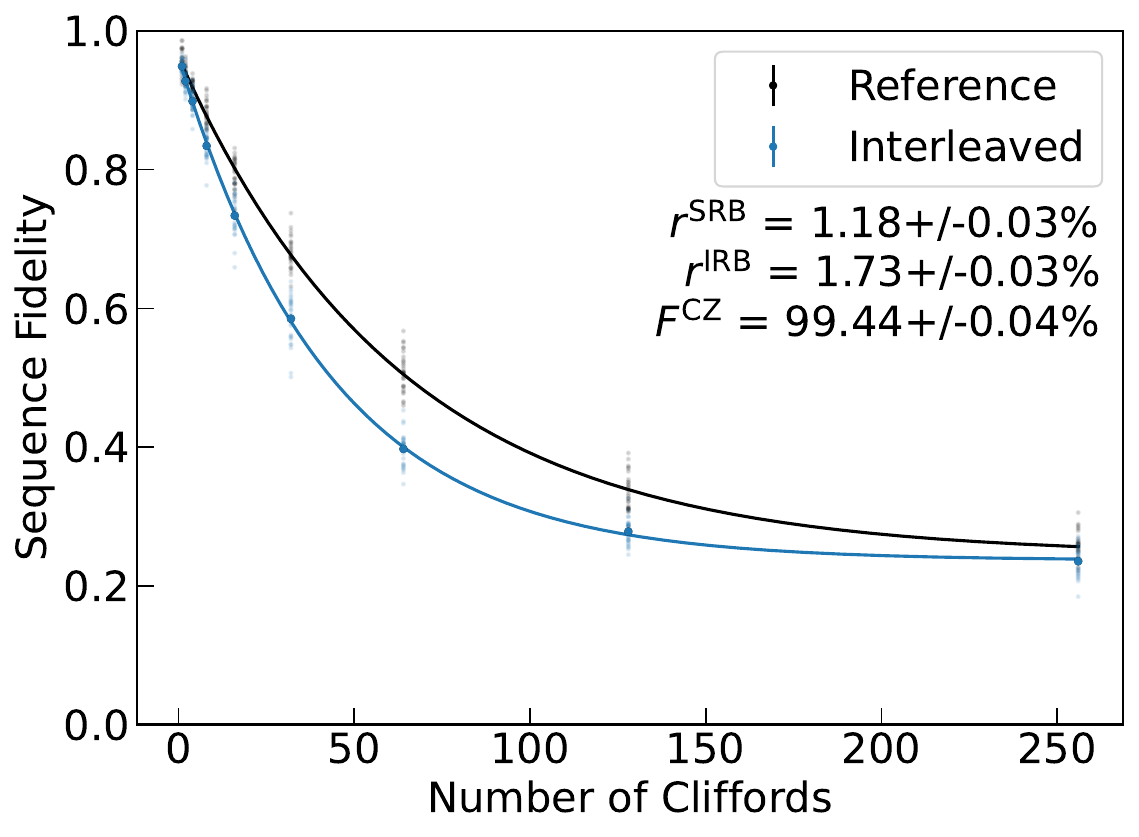}
    \caption{Adiabatic CZ gate implemented using a bi-harmonic flux pulse with a gate duration of $180~\mathrm{ns}$. The pulse is applied through the bias-tee without predistortion. Here, $r^{\mathrm{SRB}}$ and $r^{\mathrm{IRB}}$ denote the average error rates per Clifford for the standard and interleaved sequences, respectively, and $F_{\mathrm{CZ}}$ denotes the extracted CZ gate fidelity. The benchmarking data are obtained using 30 random seeds.}
    \label{fig: two-qubit gate}
\end{figure}

To illustrate the use of bi-harmonic pulses (Eq.~\eqref{eq: bi-harmonic vin}) for two-qubit operations, we implement an adiabatic CZ gate. The flux-control pulses are applied through a bias tee, without predistortion, assuming the overestimation case for RC time constant with $b_2=-2b_1$. We characterize the CZ gate fidelity using standard and interleaved randomized benchmarking (SRB and IRB, respectively)~\cite{magesan2012characterizing, magesan2012efficient, li2024realization}, obtaining $F_{\mathrm{CZ}} = 99.44(4)\,\%$ for a $180~\mathrm{ns}$ gate duration, as shown in Fig.~\ref{fig: two-qubit gate}. The corresponding qubit parameters at the idling flux bias point are listed in Table~\ref{table:device_param_2qb_gate}.

%The relatively long gate duration is attributed to the symmetric SQUID transmon coupler, which provides only a weak ZZ interaction when detuned away from the qubit frequencies, thereby reducing the contribution of the negative part of the bi-harmonic pulse to CZ phase accumulation.

\begin{table}[htbp]
\caption{Qubit parameters for Q1 and Q2 at the idling flux bias point, $\Phi \approx 0.3\,\Phi_0$, used for the CZ gate implementation, measured during a separate cryostat cooldown.}
\label{table:device_param_2qb_gate}
\begin{ruledtabular}
\renewcommand{\arraystretch}{1.3} 
\begin{tabular}{llcc}
\multicolumn{2}{l}{\textbf{Parameter}} & \textbf{Q1} & \textbf{Q2} \\ 
\hline
Qubit frequency & $\omega_q/2\pi\, (\mathrm{GHz})$ & 4.7556 & 4.6595 \\
Qubit anharmonicity & $\alpha/2\pi\, (\mathrm{MHz})$ & -240 & -240 \\
Resonator frequency & $\omega_r/2\pi \,(\mathrm{GHz})$ & 6.9975 & 6.8449 \\
Resonator linewidth & $\kappa_r/2\pi\,(\mathrm{kHz})$ & 409 & 527 \\
Relaxation time & ${T}_1 \,(\mu\mathrm{s})$ & 103 & 121 \\
Decoherence time & ${T}^*_2 \,(\mu\mathrm{s})$ & 109 & 106 \\
\hline
Coupler freq.\@ $(\Phi=0)$ & $\omega_c /2\pi\, (\mathrm{GHz})$ & 4.8575 & --- \\
Qubit-coupler coupling & $g/2\pi\, (\mathrm{MHz})$ & 77 & 80 \\
\end{tabular}
\end{ruledtabular}
\end{table}

\section{Transient Analysis for multi-pole RC ladder}
\label{app:multipole}
\begin{figure}[htbp]
    \centering
    \includegraphics[scale=0.4]{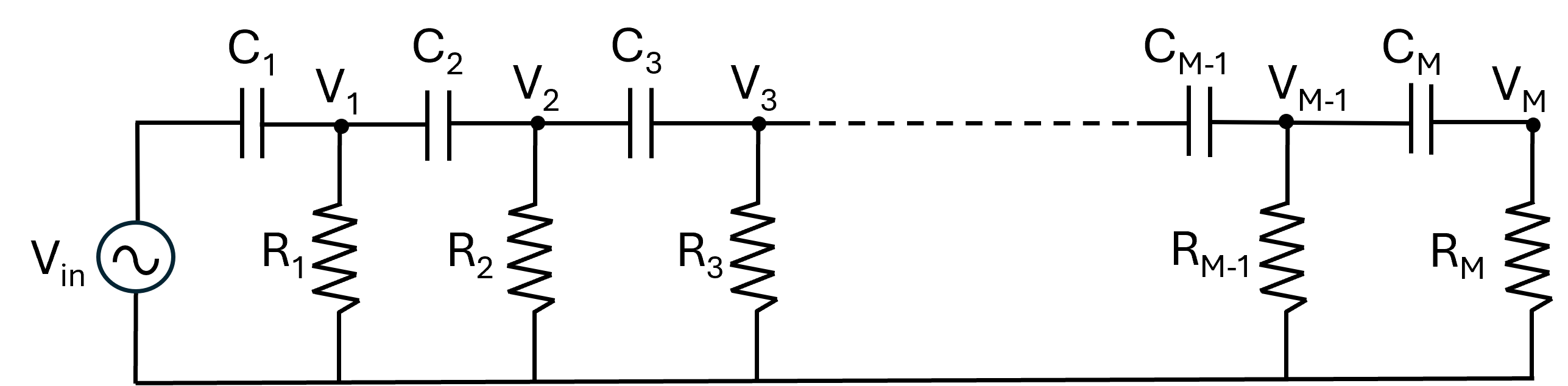}
    \caption{Multi-pole high pass RC ladder.}
    \label{fig: multi-pole}
\end{figure}

We extend the single-pole RC model to an M-stage high-pass RC ladder, consisting of $M$ capacitors $C_1,...,C_\text{M}$ connected in series along the signal path, with resistors $R_1, ...,R_\text{M}$ shunted to ground at the corresponding nodes. 
Let $V_{\mathrm{in}}(t)$ denote the input voltage and $V_\text{i}(t)$, $i=1,\ldots,M$, the voltage at the $i$-th node. Applying Kirchhoff's current law at each node gives the node transfer functions
\begin{equation}
H_\text{i}(s)\equiv\frac{V_\text{i}(s)}{V_{\text{in}}(s)}
=\frac{N_\text{i}(s)}{D_\text{M}(s)},
\qquad i=1,\ldots,M,
\label{eq}
\end{equation}
where $D_\text{M}(s)$ is a common $M$th-order denominator  shared by all node transfer functions. Denoting its poles by $s_1,\ldots,s_\text{M}$, the denominator takes the form
\begin{equation}
D_\text{M}(s)\propto\prod_{i=1}^{M}(s-s_i).
\label{eq}
\end{equation}

The quantity of interest is the current through the final resistor $R_M$, given by 
\begin{equation}
    I_\text{M}(t)=V_\text{M}(t)/R_\text{M}.
\end{equation}
Consequently, the current has the same poles as $V_\text{M}(t)$. The transient-cancellation condition can be generalized to the multi-pole ladder by requiring the residual contribution from each
pole to vanish. Defining the positive decay rates
$c_\text{i}= -s_\text{i}>0$, the condition associated with the $i$-th pole is obtained by replacing $1/\tau$ in Eq.~\eqref{eq:kexp_Laplace} with $c_\text{i}$. Thus, for each pole,
\begin{equation}
    k_{i}
    =
    \int_0^{\tau_{\mathrm{pulse}}}
    e^{c_\text{i} t'}V_{\text{in}}(t')\,dt'
    =0,
    \qquad i=1,\ldots,M.
    \label{eq:kexp_multi}
\end{equation}

Considering the Fourier-series pulse in Eq.~\eqref{eq:vin_input}, this
condition becomes
\begin{equation}
    a_0+\sum_{n=1}^{N}
    \frac{c_\text{i} a_n-n\omega b_n}
    {c_i^2+(n\omega)^2}
    =0,
    \qquad i=1,\ldots,M.
    \label{eq:M_pole_conditions_general}
\end{equation}

\section{Fitting method for transients}
In Fig.~\ref{fig:transients method}(h), we fit the flux transients curve using a fitting function with two exponential terms:
\begin{equation}
    A(-e^{-\tau_{\text{delay}}/\tau}+e^{-(\tau_{\text{delay}}+\tau_{\text{pulse}})/\tau})+B.
    \label{eq: two exp}
\end{equation}
The origin of Eq.(\ref{eq: two exp}) is due to the fact that a square-shaped pulse $x(t)$ of duration $t_0$ can be represented as 
\begin{equation}
x(t)=x_0\left(\theta(t)-\theta(t-t_0)\right),
\end{equation}
where $\theta(t)$ is the Heaviside theta function, and $x_0$ is its amplitude. To model the high-pass filter response of the flux signal, we need to convolute $x(t)$ with the  $h(t)=e^{-t/\tau} \theta(t)$, resulting in a transient response modeled by two exponential terms.

\section{Phase histogram data}
\label{app:phase_histogram}
In Table~\ref{Table: Hist}, we report the extracted mean and standard deviation of the histograms in Fig.~\ref{fig:robusteness}(d) showing that the mean values for $r = 1, 10, 100$ lie within the standard deviation of the zero (reference) case. 

\begin{table}[htbp]
\caption{\label{Table: Hist} Mean and standard deviation of the histograms in Fig.~\ref{fig:robusteness}(d).}
\begin{ruledtabular}
\begin{tabular}{ccc}
\textbf{Type} & \textbf{Mean} & \textbf{Standard deviation} \\
\hline
$r=1$   & $-0.006 \pm 0.006$ & $0.329 \pm 0.004$ \\
$r=10$  & $0.002 \pm 0.011$  & $0.618 \pm 0.008$ \\
$r=100$ & $-0.021 \pm 0.005$ & $0.254 \pm 0.003$ \\
Zero    & $-0.098 \pm 0.004$ & $0.231 \pm 0.003$ \\
\end{tabular}
\end{ruledtabular}
\end{table}

\section{Over-estimation of RC-constant and its consequences on $k_{\text{exp}}$ }
\label{appendix:under_over_estimation}

Here, we discuss details not included in the main text, but necessary to follow the derivations used for the discussion on mischaracterization of the RC constant $\tau$. We begin by solving $k_\text{exp}=0$, Eq.\eqref{eq: k_exp_condition}, where $r$ is included to account for the error in estimating $\tau$. Similarly to the main text, we consider solutions to the sine-only ($a_n = 0, \forall n$) and cosine-only ($b_n = 0 , \forall n$)  case, although it is straightforward to extend all derivations to pulses containing both terms. Note  that all results from the sine-only case can be extended to the cosine-only case with the substitution $a_n \rightarrow - r \omega \tau b_n$.

By solving Eq.\eqref{eq: k_exp_condition}, we obtain an expression for the $N$-th coefficient ($a_N$ or $b_N$, depending on the case) that would satisfy $k_\text{exp}=0$ if the RC-constant was $r\, \tau$: 
\begin{align}
    a_N^r &= - \mleft[ 1 + \mleft( N r \omega \tau \mright)^2 \mright] \sum_{n=1}^{N-1} \frac{a_n}{1 + \mleft( n r \omega \tau \mright)^2} , \\
    b_N^r &= - \frac{1 + \mleft( N r \omega \tau \mright)^2 }{N} \sum_{n=1}^{N-1} \frac{n b_n}{1 + \mleft( n r \omega \tau \mright)^2} ,
\end{align}
where the $r$ superindexes indicate that they are computed with the estimation factor $r$.

Now, we substitute these solutions in the transient coefficient $k_{\text{exp}}$. The resulting transient coefficients for the cosine- and sine-only cases, respectively, are
\begin{widetext} 
\begin{alignat}{3}
    k_{\text{exp}}^{\cos} &=& &\sum_{n=1}^{N-1} \mleft( \frac{1}{1 + \mleft( n \omega \tau \mright)^2} - \frac{1 + \mleft( N r \omega \tau \mright)^2}{\mleft[ 1 + \mleft( N \omega \tau \mright)^2 \mright] \mleft[ 1 + \mleft( n r \omega \tau \mright)^2 \mright] } \mright) a_n , \\
    k_{\text{exp}}^{\sin} &=& - &\sum_{n=1}^{N-1} \mleft( \frac{1}{1 + \mleft( n \omega \tau \mright)^2} - \frac{1 + \mleft( N r \omega \tau \mright)^2}{\mleft[ 1 + \mleft( N \omega \tau \mright)^2 \mright] \mleft[ 1 + \mleft( n r \omega \tau \mright)^2 \mright] } \mright) n \omega \tau b_n \, .
\end{alignat}

To understand the result of overestimating $\tau$, we expand these expressions as a series in $1/r$ around $1/r\rightarrow0$ (equivalently, $r \rightarrow \infty$ limit). We keep the first two terms in the series, and therefore we expect this expansion to be only accurate when the RC constant is highly overestimated:

\begin{alignat}{5}
   k_{\text{exp}}^{\cos,\gg} &=& &\sum_{n=1}^{N-1} \mleft[ \frac{1}{1 + (n \omega \tau)^2} - \frac{N^2}{n^2 (1 + (N \omega \tau)^2)} \mright] a_n + \sum_{n=1}^{N-1} \frac{N^2 - n^2}{n^4 (\omega \tau)^2 (1 + (N \omega \tau)^2)} \frac{a_n}{r^2} + \mathcal{O} \mleft( \frac{1}{r^4} \mright) ,\label{eq:cosine_gg} \\
   k_{\text{exp}}^{\sin,\gg} &=&-&\sum_{n=1}^{N-1} \mleft[ \frac{1}{1 + (n \omega \tau)^2} - \frac{N^2}{n^2 (1 + (N \omega \tau)^2)} \mright] n \omega \tau b_n - \sum_{n=1}^{N-1} \frac{(N^2 - n^2) n \omega \tau}{n^4 (\omega\tau)^2 (1 + (N \omega \tau)^2)} \frac{b_n}{r^2} + \mathcal{O} \mleft( \frac{1}{r^4} \mright),\label{eq:sine_gg}
\end{alignat}
\end{widetext}
with the $\gg$ superindex denoting $r \gg 1$. The transient coefficients are very similar to each other, but the asymmetry between $a_n$ and $b_n$ changes the scaling with $\omega \tau$.

In particular, we are interested in the regime where flux pulses are fast compared to the RC decay, i.e., $\omega \tau \gg 1$. To identify the leading-order scaling in this regime, we further expand Eqs. \eqref{eq:cosine_gg} and \eqref{eq:sine_gg} in powers of $1/(\omega\tau)$, keeping only the leading term:
 
\begin{align}
    \left. k_{\text{exp}}^{\cos} \right\rvert_{r\gg1,\,\omega \tau \gg 1} &\approx \sum_{n=1}^{N-1} \left( \frac{1}{N^2 n^2} - \frac{1}{n^4} \right) \frac{a_n}{(\omega \tau)^4} ,\label{eq:k_cos_gt} \\
    \left.k_{\text{exp}}^{\sin} \right\rvert_{r\gg1,\,\omega \tau \gg 1} &\approx \sum_{n=1}^{N-1} \left( \frac{1}{n^3} - \frac{1}{N^2 n} \right) \frac{b_n}{(\omega \tau)^3} .\label{eq:k_sin_gt} 
\end{align}
 Equations (\ref{eq:k_cos_gt}) and (\ref{eq:k_sin_gt}) show that if $\tau$ is overestimated and  $\omega \tau \gg 1$, $k_{\text{exp}}$ is further suppressed with two extra negative powers of $\omega \tau$  in comparison to Eq.~\eqref{eq: k_exp_condition}.

\bibliography{apssamp}% Produces the bibliography via BibTeX.

@article{rol2020time,
    author = {Rol, M. A. and Ciorciaro, L. and Malinowski, F. K. and Tarasinski, B. M. and Sagastizabal, R. E. and Bultink, C. C. and Salathe, Y. and Haandb{\ae}k, N. and Sedivy, J. and DiCarlo, L.},
    title = {Time-domain characterization and correction of on-chip distortion of control pulses in a quantum processor},
    journal = {Applied Physics Letters},
    volume = {116},
    pages = {054001},
    year = {2020},
    doi = {10.1063/1.5133894}
}

@article{sung2021realization,
  title = {Realization of High-Fidelity {CZ} and {$ZZ$}-Free {iSWAP} Gates with a Tunable Coupler},
  author = {Sung, Youngkyu and Ding, Leon and Braum{\"u}ller, Jochen and Veps{\"a}l{\"a}inen, Antti and Kannan, Bharath and Kjaergaard, Morten and Greene, Ami and Samach, Gabriel O. and McNally, Chris and Kim, David and others},
  journal = {Physical Review X},
  volume = {11},
  pages = {021058},
  year = {2021},
  doi = {10.1103/PhysRevX.11.021058}
}

@article{li2024realization,
  title = {Realization of High-Fidelity {CZ} Gate Based on a Double-Transmon Coupler},
  author = {Li, Rui and Kubo, Kentaro and Ho, Yinghao and Yan, Zhiguang and Nakamura, Yasunobu and Goto, Hayato},
  journal = {Physical Review X},
  volume = {14},
  pages = {041050},
  year = {2024},
  doi = {10.1103/PhysRevX.14.041050}
}

@article{glaser2024sensitivity,
  title = {Closed-loop optimization for high-fidelity controlled-{$Z$} gates in superconducting qubits},
  author = {Glaser, N. J. and Roy, F. A. and Tsitsilin, I. and Koch, L. and Bruckmoser, N. and Schirk, J. and Romeiro, J. H. and Huber, G. B. P. and Wallner, F. and Singh, M. and others},
  journal = {Physical Review Applied},
  volume = {24},
  pages = {024048},
  year = {2025},
  doi = {10.1103/pckq-2csc}
}

@phdthesis{heinsoo2019digital,
  title={Digital quantum computation with superconducting qubits},
  author={Heinsoo, Johannes},
  year={2019},
  school={ETH Zurich}
}

@article{zhang2023characterization,
  title = {Characterization of tunable coupler without a dedicated readout resonator in superconducting circuits},
  author = {Zhang, Chi and Wang, Tian-Le and Guo, Liang-Liang and Yang, Xiao-Yan and Yang, Xin-Xin and Duan, Peng and Jia, Zhi-Long and Kong, Wei-Cheng and Guo, Guo-Ping},
  journal = {Applied Physics Letters},
  volume = {122},
  pages = {024001},
  year = {2023},
  doi = {10.1063/5.0135219}
}

@article{kosen2022building,
  title={Building blocks of a flip-chip integrated superconducting quantum processor},
  author={Kosen, Sandoko and Li, Hang-Xi and Rommel, Marcus and Shiri, Daryoush and Warren, Christopher and Gr{\"o}nberg, Leif and Salonen, Jaakko and Abad, Tahereh and Bizn{\'a}rov{\'a}, Janka and Caputo, Marco and others},
  journal={Quantum Science and Technology},
  volume={7},
  pages={035018},
  year={2022},
  doi={10.1088/2058-9565/ac734b}
}

@article{li2024high,
  title={High-precision pulse calibration of tunable couplers for high-fidelity two-qubit gates in superconducting quantum processors},
  author={Li, Tian-Ming and Zhang, Jia-Chi and Chen, Bing-Jie and Huang, Kaixuan and Liu, Hao-Tian and Xiao, Yong-Xi and Deng, Cheng-Lin and Liang, Gui-Han and Chen, Chi-Tong and Liu, Yu and others},
  journal={Physical Review Applied},
  volume={23},
  pages={024059},
  year={2025},
  doi={10.1103/PhysRevApplied.23.024059}
}

@phdthesis{baur2012realizing,
  title={Realizing quantum gates and algorithms with three superconducting qubits},
  author={Baur, Matthias},
  year={2012},
  school={ETH Zurich}
}

@mastersthesis{butscher_master,
    author = {Jonas Butscher},
    title = {Shaping of Fast Flux Pulses for Two-Qubit Gates: Inverse Filtering},
    school ={ETH Z{\"u}rich} ,
    year ={2018} 
}

@article{krinner2022realizing,
  title={Realizing repeated quantum error correction in a distance-three surface code},
  author={Krinner, Sebastian and Lacroix, Nathan and Remm, Ants and Di Paolo, Agustin and Genois, Elie and Leroux, Catherine and Hellings, Christoph and Lazar, Stefania and Swiadek, Francois and Herrmann, Johannes and others},
  journal={Nature},
  volume={605},
  pages={669--674},
  year={2022},
  doi={10.1038/s41586-022-04566-8}
}

@phdthesis{johnson2011controlling,
    author = {Johnson, Blake Robert},
    title ={Controlling photons in superconducting electrical circuits} ,
    school = {Yale University},
    year = {2011} 
}

@article{rol2019fast,
  title = {Fast, High-Fidelity Conditional-Phase Gate Exploiting Leakage Interference in Weakly Anharmonic Superconducting Qubits},
  author = {Rol, M. A. and Battistel, F. and Malinowski, F. K. and Bultink, C. C. and Tarasinski, B. M. and Vollmer, R. and Haider, N. and Muthusubramanian, N. and Bruno, A. and Terhal, B. M. and others},
  journal = {Physical Review Letters},
  volume = {123},
  pages = {120502},
  year = {2019},
  doi = {10.1103/PhysRevLett.123.120502}
}

@article{chen2024fast,
  title={Fast, unconditional reset and leakage reduction in fixed-frequency transmon qubits},
  author={Chen, Liangyu and Fors, Simon Pettersson and Yan, Zixian and Ali, Anaida and Abad, Tahereh and Huang, Tangyou and Osman, Amr and Moschandreou, Eleftherios and Lienhard, Benjamin and Kosen, Sandoko and others},
  journal={npj Quantum Information},
  volume={12},
  pages={135},
  year={2026},
  doi={10.1038/s41534-026-01356-2}
}

@article{foxen2020demonstrating,
  title={Demonstrating a continuous set of two-qubit gates for near-term quantum algorithms},
  author={Foxen, Brooks and Neill, Charles and Dunsworth, Andrew and Roushan, Pedram and Chiaro, Ben and Megrant, Anthony and Kelly, Julian and Chen, Zijun and Satzinger, Kevin and Barends, Rami and others},
  journal={Physical Review Letters},
  volume={125},
  pages={120504},
  year={2020},
  doi={10.1103/PhysRevLett.125.120504}
}

@article{marxer2023long,
  title={Long-distance transmon coupler with {CZ}-gate fidelity above 99.8\%},
  author={Marxer, Fabian and Veps{\"a}l{\"a}inen, Antti and Jolin, Shan W. and Tuorila, Jani and Landra, Alessandro and Ockeloen-Korppi, Caspar and Liu, Wei and Ahonen, Olli and Auer, Adrian and Belzane, Lucien and others},
  journal={PRX Quantum},
  volume={4},
  pages={010314},
  year={2023},
  doi={10.1103/PRXQuantum.4.010314}
}

@article{reed2010fast,
  title = {Fast reset and suppressing spontaneous emission of a superconducting qubit},
  author = {Reed, M. D. and Johnson, B. R. and Houck, A. A. and DiCarlo, L. and Chow, J. M. and Schuster, D. I. and Frunzio, L. and Schoelkopf, R. J.},
  journal = {Applied Physics Letters},
  volume = {96},
  pages = {203110},
  year = {2010},
  doi = {10.1063/1.3435463}
}

@article{yang2024coupler,
  title={Coupler-assisted leakage reduction for scalable quantum error correction with superconducting qubits},
  author={Yang, Xiaohan and Chu, Ji and Guo, Zechen and Huang, Wenhui and Liang, Yongqi and Liu, Jiawei and Qiu, Jiawei and Sun, Xuandong and Tao, Ziyu and Zhang, Jiawei and others},
  journal={Physical Review Letters},
  volume={133},
  pages={170601},
  year={2024},
  doi={10.1103/PhysRevLett.133.170601}
}

@article{acharya2024quantum,
  title={Quantum error correction below the surface code threshold},
  author={{Google Quantum AI and Collaborators}},
  journal={Nature},
  volume={638},
  pages={920--926},
  year={2025},
  doi={10.1038/s41586-024-08449-y}
}

@article{chayanun2024characterization,
  title={Characterization of process-related interfacial dielectric loss in aluminum-on-silicon by resonator microwave measurements, materials analysis, and imaging},
  author={Chayanun, Lert and Bizn{\'a}rov{\'a}, Janka and Zeng, Lunjie and Malmberg, Per and Nylander, Andreas and Osman, Amr and Rommel, Marcus and Tam, Pui Lam and Olsson, Eva and Delsing, Per and others},
  journal={APL Quantum},
  volume={1},
  pages={026115},
  year={2024},
  doi={10.1063/5.0208140}
}

@article{rehammar2023low,
  title={Low-pass filter with ultrawide stopband for quantum computing applications},
  author={Rehammar, Robert and Gasparinetti, Simone},
  journal={IEEE Transactions on Microwave Theory and Techniques},
  volume={71},
  pages={3075--3080},
  year={2023},
  doi={10.1109/TMTT.2023.3238543}
}

@article{marques2022logical,
  title={Logical-qubit operations in an error-detecting surface code},
  author={Marques, J. F. and Varbanov, B. M. and Moreira, M. S. and Ali, H. and Muthusubramanian, N. and Zachariadis, C. and Battistel, F. and Beekman, M. and Haider, N. and Vlothuizen, W. and others},
  journal={Nature Physics},
  volume={18},
  pages={80--86},
  year={2022},
  doi={10.1038/s41567-021-01423-9}
}

@article{jurcevic2021demonstration,
  title={Demonstration of quantum volume 64 on a superconducting quantum computing system},
  author={Jurcevic, Petar and Javadi-Abhari, Ali and Bishop, Lev S. and Lauer, Isaac and Bogorin, Daniela F. and Brink, Markus and Capelluto, Lauren and G{\"u}nl{\"u}k, Oktay and Itoko, Toshinari and Kanazawa, Naoki and others},
  journal={Quantum Science and Technology},
  volume={6},
  pages={025020},
  year={2021},
  doi={10.1088/2058-9565/abe519}
}

@article{xu2020high,
  title={High-fidelity, high-scalability two-qubit gate scheme for superconducting qubits},
  author={Xu, Yuan and Chu, Ji and Yuan, Jiahao and Qiu, Jiawei and Zhou, Yuxuan and Zhang, Libo and Tan, Xinsheng and Yu, Yang and Liu, Song and Li, Jian and others},
  journal={Physical Review Letters},
  volume={125},
  pages={240503},
  year={2020},
  doi={10.1103/PhysRevLett.125.240503}
}

@article{zhao2022realization,
  title={Realization of an error-correcting surface code with superconducting qubits},
  author={Zhao, Youwei and Ye, Yangsen and Huang, He-Liang and Zhang, Yiming and Wu, Dachao and Guan, Huijie and Zhu, Qingling and Wei, Zuolin and He, Tan and Cao, Sirui and others},
  journal={Physical Review Letters},
  volume={129},
  pages={030501},
  year={2022},
  doi={10.1103/PhysRevLett.129.030501}
}

@article{mcewen2021removing,
  title={Removing leakage-induced correlated errors in superconducting quantum error correction},
  author={McEwen, M. and Kafri, D. and Chen, Z. and Atalaya, J. and Satzinger, K. J. and Quintana, C. and Klimov, P. V. and Sank, D. and Gidney, C. and Fowler, A. G. and others},
  journal={Nature Communications},
  volume={12},
  pages={1761},
  year={2021},
  doi={10.1038/s41467-021-21982-y}
}

@article{lacroix2023fast,
  title={Fast flux-activated leakage reduction for superconducting quantum circuits},
  author={Lacroix, Nathan and Hofele, Luca and Remm, Ants and Benhayoune-Khadraoui, Othmane and McDonald, Alexander and Shillito, Ross and Lazar, Stefania and Hellings, Christoph and Swiadek, Francois and Colao-Zanuz, Dante and others},
  journal={Physical Review Letters},
  volume={134},
  pages={120601},
  year={2025},
  doi={10.1103/PhysRevLett.134.120601}
}

@article{guo2024universal,
  title={Universal scalable characterization and correction of pulse distortions in controlled quantum systems},
  author={Guo, Liang-Liang and Duan, Peng and Zhang, Sheng and Yang, Xin-Xin and Zhang, Chi and Du, Lei and Zhang, Hai-Feng and Tao, Hao-Ran and Wang, Tian-Le and Jia, Zhi-Long and others},
  journal={Physical Review Applied},
  volume={21},
  pages={064060},
  year={2024},
  doi={10.1103/PhysRevApplied.21.064060}
}

@article{biznarova2024mitigation,
  title={Mitigation of interfacial dielectric loss in aluminum-on-silicon superconducting qubits},
  author={Bizn{\'a}rov{\'a}, Janka and Osman, Amr and Rehnman, Emil and Chayanun, Lert and Kri{\v{z}}an, Christian and Malmberg, Per and Rommel, Marcus and Warren, Christopher and Delsing, Per and Yurgens, August and others},
  journal={npj Quantum Information},
  volume={10},
  pages={78},
  year={2024},
  doi={10.1038/s41534-024-00868-z}
}

@article{koch2007charge,
  title={Charge-insensitive qubit design derived from the Cooper pair box},
  author={Koch, Jens and Yu, Terri M. and Gambetta, Jay and Houck, Andrew A. and Schuster, David I. and Majer, Johannes and Blais, Alexandre and Devoret, Michel H. and Girvin, Steven M. and Schoelkopf, Robert J.},
  journal={Physical Review A},
  volume={76},
  pages={042319},
  year={2007},
  doi={10.1103/PhysRevA.76.042319}
}

@book{oppenheim1997signals,
  title={Signals and Systems},
  author={Oppenheim, Alan V. and Willsky, Alan S. and Nawab, S. Hamid},
  publisher={Prentice Hall},
  year={1997}
}

@article{bengtsson2020improved,
  title={Improved success probability with greater circuit depth for the quantum approximate optimization algorithm},
  author={Bengtsson, Andreas and Vikst{\aa}l, Pontus and Warren, Christopher and Svensson, Marika and Gu, Xiu and Kockum, Anton Frisk and Krantz, Philip and Kri{\v{z}}an, Christian and Shiri, Daryoush and Svensson, Ida-Maria and others},
  journal={Physical Review Applied},
  volume={14},
  pages={034010},
  year={2020},
  doi={10.1103/PhysRevApplied.14.034010}
}

@article{mckay2016universal,
  title={Universal gate for fixed-frequency qubits via a tunable bus},
  author={McKay, David C. and Filipp, Stefan and Mezzacapo, Antonio and Magesan, Easwar and Chow, Jerry M. and Gambetta, Jay M.},
  journal={Physical Review Applied},
  volume={6},
  pages={064007},
  year={2016},
  doi={10.1103/PhysRevApplied.6.064007}
}

@misc{scipy_savgol_filter,
  author       = {SciPy},
  title        = {scipy savgol filter},
  year         = {2024},
  url          = {https://docs.scipy.org/doc/scipy/reference/generated/scipy.signal.savgol_filter.html}
}

@article{jerger2019situ,
  title={In situ characterization of qubit control lines: A qubit as a vector network analyzer},
  author={Jerger, Markus and Kulikov, Anatoly and Vasselin, Z{\'e}non and Fedorov, Arkady},
  journal={Physical Review Letters},
  volume={123},
  pages={150501},
  year={2019},
  doi={10.1103/PhysRevLett.123.150501}
}

@article{zhang2025characterization,
  title={Characterization and optimization of tunable couplers via adiabatic control in superconducting circuits},
  author={Zhang, Xuan and Zhang, Xu and Chen, Changling and Tang, Kai and Yi, Kangyuan and Luo, Kai and Xie, Zheshu and Chen, Yuanzhen and Yan, Tongxing},
  journal={Physical Review Applied},
  volume={24},
  pages={034003},
  year={2025},
  doi={10.1103/wgqf-8rcy}
}

@article{foxen2018high,
  title={High speed flux sampling for tunable superconducting qubits with an embedded cryogenic transducer},
  author={Foxen, B. and Mutus, J. Y. and Lucero, E. and Jeffrey, E. and Sank, D. and Barends, R. and Arya, K. and Burkett, B. and Chen, Yu and Chen, Zijun and others},
  journal={Superconductor Science and Technology},
  volume={32},
  pages={015012},
  year={2019},
  doi={10.1088/1361-6668/aaf048}
}

@article{wenner2011wirebond,
  title={Wirebond crosstalk and cavity modes in large chip mounts for superconducting qubits},
  author={Wenner, James and Neeley, Matthew and Bialczak, Radoslaw C. and Lenander, Michael and Lucero, Erik and O'Connell, Aaron D. and Sank, Daniel and Wang, Haohua and Weides, Martin and Cleland, Andrew N. and others},
  journal={Superconductor Science and Technology},
  volume={24},
  pages={065001},
  year={2011},
  doi={10.1088/0953-2048/24/6/065001}
}

@phdthesis{li2025flip,
  title={Flip-chip Integrated Superconducting Quantum Processors},
  author={Li, Hang-Xi},
  school={Chalmers University of Technology}, 
  year={2025}
}

@book{footnote1,
  title = {\textnormal{Since our model is a series RC circuit, any transients in the voltage across the capacitor ($ V_c $) translate into transients in the current as $ I = C \frac{d V_c(t)}{dt}$.}},
}

@dataset{aggarwal_2025_transients_flux,
  author       = {Aggarwal, Anuj},
  title        = {Dataset and code used for Mitigating transients in flux-control signals in a superconducting quantum processor},
  year         = {2025},
  publisher    = {Zenodo},
  version      = {1.0},
  doi          = {10.5281/zenodo.17177282},
  url          = {https://doi.org/10.5281/zenodo.17177282}
}

@article{an2026zz,
  title={{ZZ}-free two-transmon {CZ} gate mediated by a fluxonium coupler},
  author={An, Junyoung and Zhang, Helin and Ding, Qi and Ding, Leon and Sung, Youngkyu and Winik, Roni and Kim, Junghyun and Rosen, Ilan T. and Azar, Kate and Pi{\~n}ero, Ren{\'e}e DePencier and others},
  journal={Physical Review Applied},
  volume={26},
  pages={014076},
  year={2026},
  doi={10.1103/9k19-c99g}
}

@article{magesan2012characterizing,
  title={Characterizing quantum gates via randomized benchmarking},
  author={Magesan, Easwar and Gambetta, Jay M. and Emerson, Joseph},
  journal={Physical Review A},
  volume={85},
  pages={042311},
  year={2012},
  doi={10.1103/PhysRevA.85.042311}
}

@article{magesan2012efficient,
  title={Efficient measurement of quantum gate error by interleaved randomized benchmarking},
  author={Magesan, Easwar and Gambetta, Jay M. and Johnson, Blake R. and Ryan, Colm A. and Chow, Jerry M. and Merkel, Seth T. and Da Silva, Marcus P. and Keefe, George A. and Rothwell, Mary B. and Ohki, Thomas A. and others},
  journal={Physical Review Letters},
  volume={109},
  pages={080505},
  year={2012},
  doi={10.1103/PhysRevLett.109.080505}
}

@article{hellings2025calibrating,
  title={Calibrating magnetic flux control in superconducting circuits by compensating distortions on timescales from nanoseconds up to tens of microseconds},
  author={Hellings, Christoph and Lacroix, Nathan and Remm, Ants and Boell, Richard and Herrmann, Johannes and Laz{\u{a}}r, Stefania and Krinner, Sebastian and Swiadek, Fran{\c{c}}ois and Andersen, Christian Kraglund and Eichler, Christopher and others},
  journal={Physical Review Research},
  volume={7},
  pages={043142},
  year={2025},
  doi={10.1103/1qhb-r4fb}
}

@article{barends2014superconducting,
  title={Superconducting quantum circuits at the surface code threshold for fault tolerance},
  author={Barends, Rami and Kelly, Julian and Megrant, Anthony and Veitia, Andrzej and Sank, Daniel and Jeffrey, Evan and White, Ted C. and Mutus, Josh and Fowler, Austin G. and Campbell, Brooks and others},
  journal={Nature},
  volume={508},
  pages={500--503},
  year={2014},
  doi={10.1038/nature13171}
}

@PREAMBLE{
 "\providecommand{\noopsort}[1]{}" 
 # "\providecommand{\singleletter}[1]{#1}%" 
}

\end{document}